\documentclass[amsmath,amssymb,aps,floatfix,onecolumn,notitlepage]{revtex4-1}

\newcommand{\brancha}{a}
\newcommand{\branchb}{b} 

\usepackage{amsmath}
\usepackage{graphicx}
\usepackage{bm}
\usepackage{nicefrac}
\usepackage{xcolor}

\DeclareMathOperator{\e}{\mathrm{e}}
\let\d=\relax
\DeclareMathOperator{\d}{\mathrm{d} \!}
\newcommand{\new}[1]{{#1}}
\renewcommand{\emph}[1]{\textit{#1}}

\begin{document}
	
	
	\title{Branch length statistics in phylogenetic trees under constant-rate birth-death dynamics}
	
	\author{Tobias Dieselhorst and Johannes Berg}
	\email{bergj@uni-koeln.de}
	\affiliation{Institute for Biological Physics, University of Cologne,
		Z\"ulpicher Stra{\ss}e 77a, 50937 Cologne, Germany}
	
	\date{\today}
	
	\begin{abstract}
		Phylogenetic trees represent the evolutionary relationships between extant lineages, where extinct or non-sampled lineages are omitted. Extending the work of Stadler and collaborators, this paper focuses on the branch lengths in phylogenetic trees arising under a constant-rate birth-death model. We derive branch length distributions of phylogenetic branches with and without random sampling of individuals of the extant population under two distinct statistical scenarios: a fixed age of the birth-death process and a fixed number of individuals at the time of observation. We find that branches connected to the tree leaves (pendant branches) and branches in the interior of the tree behave very differently under sampling; pendant branches grow longer without limit as the sampling probability is decreased, whereas the interior branch lengths quickly reach an asymptotic distribution that does not depend on the sampling probability. 
	\end{abstract}
	
	\maketitle
	
	\tableofcontents

	\section{Introduction}
	
	The reconstructed evolutionary process is central to understanding evolutionary relationships and dynamics. 
	This reconstruction is based on extant genetic data and yields a phylogenetic tree which represents the evolutionary relationships among a set of species, genes, or cells. Crucially, all lineages that eventually went extinct (or were not sampled) are missing from such a tree. This affects the statistics of the reconstructed tree: Under a stochastic birth-death process, species go extinct at a specific rate (or cells die, in a cellular system). Looking at the reconstructed tree, however, only speciation events are seen. To compare empirical phylogenetic trees with models of evolution, it is crucial to understand the link between 
	the underlying evolutionary process and the reconstructed process. 
	
	The reconstructed evolutionary process has been studied extensively, with early work of Nee, May, and Harvey~\cite{nee_reconstructed_1997}
	giving the probability density in time for each speciation event in a reconstructed tree from a birth-death process with 
	constant rates. 
	Such a birth-death model describes a population of species, individuals, or cells dividing independently at a fixed rate $\lambda$ and dying independently at a fixed rate $\mu$. For species, birth events correspond to speciations, whereas death events correspond to extinctions. A constant-rate birth-death model serves as a simple model of neutral evolution. 
	
	Stadler and collaborators have extended the work by Nee, May, and Harvey~\cite{nee_reconstructed_1997} to trees reconstructed from individuals that were sampled from a population~\cite{stadler_incomplete_2009} and to populations conditioned on a particular size~\cite{gernhard_conditioned_2008}. The resulting statistics has been used to infer trees and their characteristics~\cite{stadler_recovering_2013,macpherson_unifying_2022}, and is the 
	cornerstone of widely-used computational packages such as BEAST~\cite{bouckaert_beast_2019} and TreeTime~\cite{sagulenko_treetime_2018}. Many extensions of the basic constant-rate birth-death process have been developed, for instance, to account for rates that are variable through time or vary across the tree, see~\cite{stadler_recovering_2013} for an overview. 
	
	In~\cite{stadler_distribution_2012,mooers_branch_2012} Stadler, Steel and collaborators focus on a particular aspect of tree statistics, namely the distribution of branch lengths in calendar time. They calculate the probability density function for the length of the so-called pendant branches (the lowest branches leading to the leaves of the tree) for trees reconstructed from a given number of leaves. For the particular case of a Yule process (a birth-death process with the rate of death set to zero), they also give the probability density function for the length of interior branches. In~\cite{paradis_distribution_2016}, Paradis introduces an approach to calculate branch length distributions of both pendant and interior branches 
	at a particular time since the start of the birth-death process. 
	
	In this paper, we revisit the problem of the distribution of branch lengths with a particular focus on the 
	trees reconstructed from a finite fraction of extant individuals sampled at the time of observation.
	\new{We find that when the sampled fraction of the population is lowered, pendant branches increase in length while interior branches converge to an asymptotic distribution which is independent of the sampling fraction.}
	\new{In our analysis,} we distinguish two distinct statistical scenarios: In the first scenario (scenario i)), the time interval between the start of the birth-death process with a single individual and the time of observation (the end of the birth-death process) is known, but the number of extant individuals at the end of the process is a fluctuating random variable. In the second scenario (scenario ii)), one conditions on the 
	number of individuals at the end of the process, but the time interval between the start and end of the process is unknown.
	
	In Section~\ref{sec:birth_death}, we recapitulate the relevant properties of a constant-rate birth-death process. In Sections~\ref{sec:branches_pendant}-\ref{sec:branches_interior} we use the approach introduced by Paradis~\cite{paradis_distribution_2016} to systematically write down concrete expressions for branch length distributions for different types of branches under the first scenario: In Section~\ref{sec:branches_pendant} we focus on pendant branches, in Section~\ref{sec:branches_interior_above_pendant} on branches one and two levels above a pendant branch, and in Section~\ref{sec:branches_interior} on generic branches in the interior of a phylogenetic tree. Sections~\ref{sec:branches_pendant} and~\ref{sec:branches_interior} coincide with the work of Paradis in~\cite{paradis_distribution_2016}; they are included here for completeness and because~\cite{paradis_distribution_2016} does not give concrete expressions for the branch length distributions.     
	Section~\ref{sec:sampling} gives the key results of this paper; we consider branch lengths arising in trees reconstructed from only a finite fraction of the extant population. We find that sampling affects different parts of the reconstructed tree differently: while pendant branches become longer and grow without limit as the sampling probability is taken to zero, the distribution of interior branch lengths quickly reaches an asymptotic limit as the sampling probability decreases. In Section~\ref{sec:scenario2} we turn to the second scenario (conditioning on the number of extant individuals at the time of observation) and introduce a novel approach based on tracking the birth-death process first backwards in time from a given number of individuals at the time of observation to the emergence of a particular branch and then again forwards again to the time of observation. For pendant branches, we recover the results of Stadler, Steel and collaborators~\cite{stadler_distribution_2012,mooers_branch_2012} and generalize these results to interior branches. 
	
	\section{Dynamics of a birth-death process}
	\label{sec:birth_death}
	
	We consider a standard birth-death process where individuals in a population duplicate with a constant rate of birth $\lambda$ and die with a constant rate of death $\mu$. These individuals might describe individual cells (for instance bacteria or tumour cells) undergoing cell division and cell death, individual multi-cellular organisms in a population under asexual reproduction, or species undergoing speciation and extinction events. Throughout, we will refer to individuals undergoing birth and death events for consistency. The dynamics of the number of individuals $n$ is described by a Markov process, with the probability $p_n(t)$ of the population consisting of $n$ individuals at time $t$ following the master equation
	\begin{equation}
		\label{eq:master_pn}
		\partial_t p_n(t) = \lambda (n-1) p_{n-1}(t)-\lambda n p_{n}(t) + \mu (n+1) p_{n+1}(t) - \mu n p_{n}(t) \ .
	\end{equation}

	Starting with a single individual at time $t=0$ and with a birth rate larger than the death rate $\lambda>\mu$, a population can grow exponentially with time, or it can die out at some time. 
	Kendall \cite{kendall_generalized_1948} derived the probability that such a population has died out by the time $t>0$ 
	\begin{equation}
		\label{eq:p0}
		p_0(t)= \mu \frac{1-\e^{-(\lambda-\mu)t}}{\lambda-\mu\e^{-(\lambda-\mu)t}}\ ,
	\end{equation}
	as well as the probability that at time $t$, there is a single living individual 
	\begin{equation}
		\label{eq:p1}
		p_1(t)=\frac{(\lambda-\mu)^2 \e^{-(\lambda-\mu)t}}{(\lambda-\mu \e^{-(\lambda-\mu)t})^2} \ ,
	\end{equation}
	and the probability that at time $t$ the population has size $n$ 
	\begin{align}
		\label{eq:pn}
		p_n(t) &= \left(1-p_0(t)\right) \left(1-\frac{\lambda}{\mu}p_0(t)\right) \left(\frac{\lambda}{\mu}p_0(t)\right)^{n-1} \ .
	\end{align}
	The expected population size grows exponentially with
	\begin{align}
		\label{eq:expect_n}
		\mathrm{E}\left[n(t)\right] = \sum_{n=0}^\infty np_n(t) = \e^{(\lambda-\mu)t} \ .
	\end{align}
	
	A powerful alternative to the master equation \eqref{eq:master_pn} is a master equation for the probability that a population that started with a single individual at time zero has population size one at time $t$~\cite{maddison_estimating_2007,stadler_sampling-through-time_2010}. A generalization of this approach is the probability that in a \new{clade} that started with a single individual at time \new{$t_0$}, at time \new{$t_e$} there is one and only one individual that has surviving offspring at a later time $T$. This probability turns out to be a key tool to calculate the distribution of branch lengths. 
	
	Specifically, $\tilde{p}^{(t_0,T)}_1(t_e)$ denotes the probability that at time $t_e$ a clade which started at time \new{$t_0$} contains one individual with open future fate, and all other individuals alive at time \new{$t_e>t_0$} have no extant descendants at time $T$ (including themselves). \new{This definition appears cumbersome at first, but is crucial to characterize a branch; all lineages that split off from the branch in bifurcations between its start \new{$t_0$} and end \new{$t_e$} must die out by the time of observation, otherwise the branch ends before \new{$t_e$}. }
	The master equation for $\tilde{p}^{(\new{t_0,}T)}_1(\new{t_e})$ is 
	\begin{equation}
		\label{eq:master_p1}
		\partial_{\new{t_e}} \tilde{p}^{(\new{t_0,}T)}_1(\new{t_e}) = - \mu \tilde{p}^{(\new{t_0,}T)}_1(\new{t_e}) - \lambda\left( 1-2 p_0(T-\new{t_e})\right)\tilde{p}^{(\new{t_0,}T)}_1(\new{t_e}) \ ,
	\end{equation}
	where the first term accounts for the decay of probability due to that individual's death, and the second term describes the probability that any additional lineage arising in a birth event fails to become extinct by the time of observation. The factor of two comes from the two offspring in such a birth event which can play this role. This is a master equation in a single probability only; the population aspect of this dynamics resides compactly in the probability $p_0(T-\new{t_e})$ that a particular clade dies out in a time interval of length $T-\new{t_e}$ given by~\eqref{eq:p0}.
	
	The master equation \eqref{eq:master_p1} is a first-order linear differential equation solved by \new{an exponential function with} the integrating factor
	$-\int_\new{t_0}^{\new{t_e}} \d t' \left[\mu+\lambda\left( 1-2 p_0(T-t')\right)\right]$ \new{as its argument}. Integrating to $\new{t_e}=T$ yields the probability of population size one at time $T$ \new{(equation \eqref{eq:p1} when $t_0=0$)}.
	Integrating $\new{t_e}$ only to some intermediate time $T-\tau_e$ gives the probability
	$\tilde{p}^{(\new{t_0,}T)}_1(T-\tau_e)$. (The name $\tau_e$ is chosen as this will later mark the end of an interior branch.)
	
\begin{figure}
		\centering
		\includegraphics[height=0.20\textwidth]{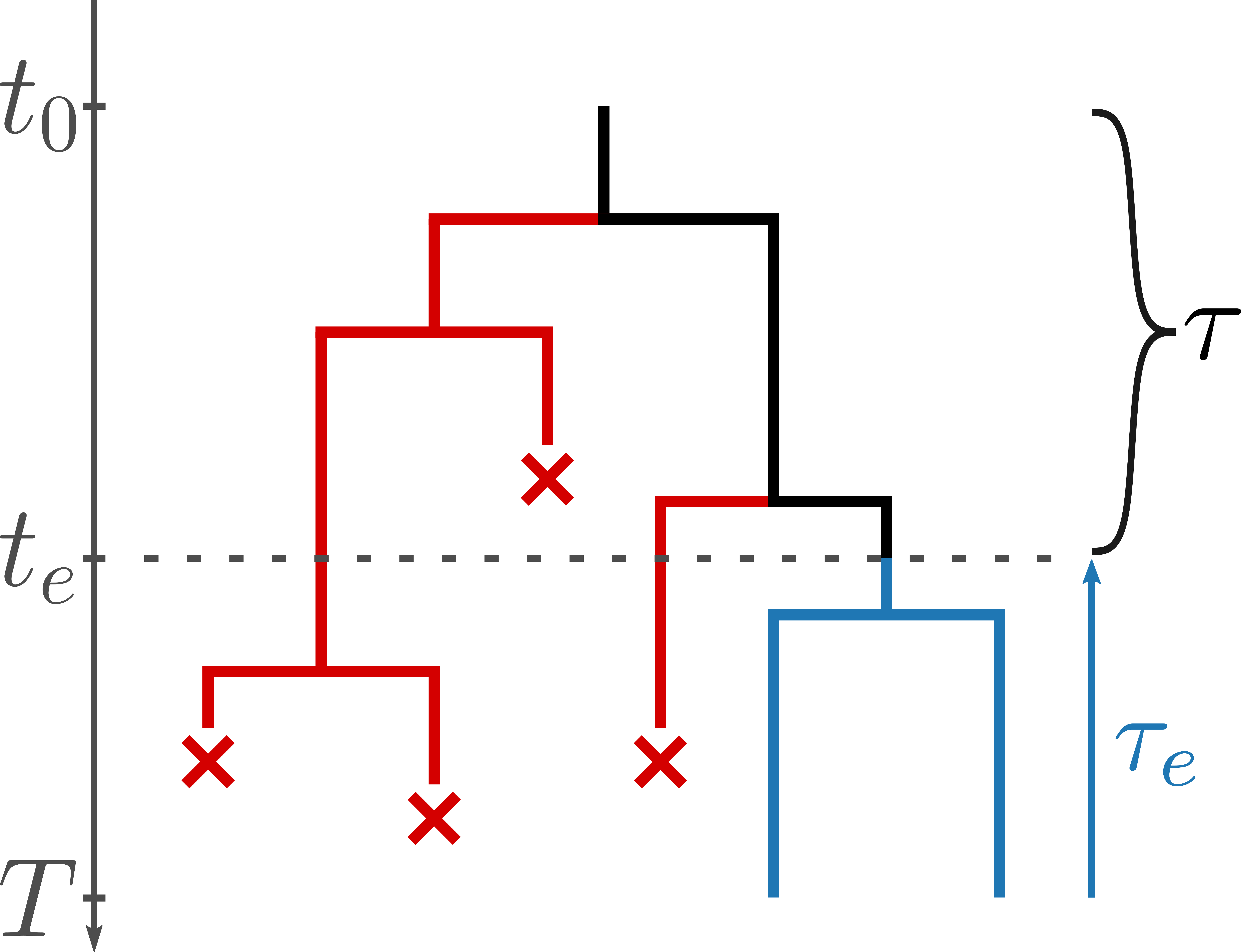}
		\caption{\textbf{Times on a birth-death tree.} 
		\new{At time $t_0$ (with time running forwards), a clade is founded by a single individual; birth and death events then change the number of individuals in the clade over time. We focus on the probability $\tilde{p}_1^{(\new{t_0,}T)}(t_e)$ that 
		the clade at time $t_e$ has the following property: all individuals and their descendants will die out by the time of observation $T$ (lineages shown in red), except for one particular individual (black lineage) whose fate is unspecified. Here this individual will later give rise to two extant individuals shown in blue. In the corresponding reconstructed phylogenetic tree, there are only the black and blue lineages. For convenience, we define $\tau_e=T-t_e$ (which runs backwards, thus increasing as we go into the evolutionary past), and $\tau$ as the interval between the origin of the clade and $\tau_e$. Using this notation, the probability defined above is written as $p_1^{(\tau_e)}(\tau)=\tilde{p}^{(t_0,T)}_1(t_e=T-\tau_e)$, see text.}  
		}\label{fig:p1_deriv}
	\end{figure}

	It turns out to be convenient for the analysis of branch lengths to consider times relative to the time of observation $T$.
	We denote the length of the interval between the intermediate time $\new{t_e}$ and the observation time $T$ by $\tau_e=T-\new{t_e}$. We further define $\tau=T\new{-t_0}-\tau_e$ as the length of the interval between $\tau_e$ and the time of the clade's origin. 
	Specifically, we denote as $p_1^{(\tau_e)}(\tau)=\tilde{p}^{(\new{t_0,}T)}_1(\new{t_e}=T-\tau_e)$ the probability that a clade started at 
	time $\tau_e+\tau$ in the past, and contains at time $\tau_e$ in the past only a single individual that could 
	have extant descendants at the time of observation \new{while all additional lineages die out before observation (see explanation above for $\tilde{p}^{(\new{t_0,}T)}_1(\new{t_e})$)}. Again, setting $\tau_e=0$ results in $p_1^{(\tau_e=0)}(\tau)\equiv p_1(\tau)$. For the case of $\tau_e>0$ the integrating factor becomes \new{$\displaystyle -\int_{t_0}^{T-\tau_e} \d t' \left[\mu+\lambda\left( 1-2 p_0(T-t')\right)\right]
	= -\int_0^{\tau} \d \tau' \left[\mu+\lambda\left( 1-2 p_0(\tau_e+\tau')\right)\right]
	=\ln\left(\frac{(\lambda-\mu\e^{-(\lambda-\mu)\tau_e})^2 \e^{-(\lambda-\mu)\tau}}{(\lambda-\mu \e^{-(\lambda-\mu)(\tau_e+\tau)})^2}\right)$} and we obtain
	\begin{equation}
		\label{eq:p1_tau1}
		p_1^{(\tau_e)}(\tau)=\frac{(\lambda-\mu\e^{-(\lambda-\mu)\tau_e})^2 \e^{-(\lambda-\mu)\tau}}{(\lambda-\mu \e^{-(\lambda-\mu)(\tau_e+\tau)})^2} \ .
	\end{equation}
	This probability differs from $p_{1}(\tau) \equiv p^{(\tau_e=0)}_{1}(\tau)$, as for $\tau_e>0$ additional lineages have more time to die out than in pendant branches
	(specifically $\tau_e$ more). This will lead to a difference between the branch length statistics of interior and pendant branches. In the following sections, we will measure times $\tau$ relative to the time of observation, with $\tau$ increasing as one looks further into the evolutionary past, see Figure~\ref{fig:p1_deriv}. 
	
	\section{Branch length distribution: pendant branches}
	\label{sec:branches_pendant}
	
	The results from section \ref{sec:birth_death} can be used directly to determine the statistics of branch lengths of a phylogenetic tree. 
	In this section, we apply them to the so-called pendant branches, which end in a leaf of the phylogenetic tree. We use the approach of Paradis~\cite{paradis_distribution_2016}
	to determine branch length distributions and write down concrete expressions for the probability densities. This section and the following are concerned with complete sampling (where the phylogenetic tree is constructed from the entire population alive at a particular time since the start of the birth-death process), in Section~\ref{sec:sampling} we will turn to the case where only a finite fraction of the population is available to reconstruct a tree from.  
	
	A pendant branch (a branch terminating at the time of observation) of length $\tau$ \new{is defined by} a birth event at time $\tau$, where (i) one of the two offspring produces a clade consisting of a single extant member at the time of observation, and (ii) the other offspring produces a clade that did not die out by the time of observation. Figure \ref{fig:interior}A shows a simple example. 
	Conditions (i) and (ii) are statistically independent events, whose probabilities can be expressed in terms of the quantities computed above. 
	This leaves the rate at which birth events occur per small time interval, which is
	given by the birth rate $\lambda$ times the population size at time $\tau$. The probability of two or more birth events occurring in the same time interval vanishes quadratically with the small time interval and can be neglected (compared to the probability of a single birth event). The population size is a random variable, and we sum over
	all possible population sizes $n$ and their probabilities, yielding the rate of a birth event at time $\tau$ as the
	expectation value of the population size multiplied with the birth rate $\lambda$. Plugging $t = T-\tau$ in equation~\eqref{eq:expect_n} gives the expected population size at time $\tau$ before sampling is \new{$\e^{(\lambda-\mu)(T-\tau)}$}.
	
	The \new{expected} number of pendant branches of length $\tau$ per small time interval is thus given by
	\begin{align}
		&\lambda\e^{(\lambda-\mu)\new{(T-\tau)}} \left[ p_1(\tau) p_{>1}(\tau) + p_{>1}(\tau)p_1(\tau) + 2 p_1(\tau)^2  \right]\\
		&= 2\lambda \e^{(\lambda-\mu)\new{(T-\tau)}} p_{1}(\tau) \left(1-p_0(\tau)\right) . 
	\end{align}
	The factor of two at the end of the first term accounts for the emergence of two branches of length $\tau$ when both offspring only have a single surviving descendant. $p_{>1}(\tau)=\sum_{n=2}^{\infty} p_n(\tau)$ denotes the probability that a clade founded at time $\tau$ has more than one extant member at the time of observation. \new{The corresponding relative frequencies of branch
	lengths give the probability distribution of $\tau$ up to a prefactor (to be determined by normalization)}.
	Using equations \eqref{eq:p0} and \eqref{eq:p1}, \new{we obtain the probability density function of branch lengths up to a normalizing factor (see below)}
	\begin{align}
		\label{eq:ptau}
		P_\text{pend}(\tau)
		&= 2 \lambda\e^{-(\lambda-\mu)\tau} p_{1}(\tau) \left(1-p_0(\tau)\right)  \\
		&= 2\lambda\frac{(\lambda-\mu)^3\e^{(\lambda-\mu)\tau}}{\left(\lambda\e^{(\lambda-\mu)\tau}-\mu\right)^3} \nonumber \ .
	\end{align}
	\new{This is the probability density function (probability per small time interval) of the lengths of pendant branches for the case of asymptotically long running times $T\to\infty$, where the normalizing factor turns out to be one for the interval $0<\tau<T=\infty$.}
	For a finite time since the start of the birth-death process, $\tau$ has a finite support and hence $P_\text{pend}(\tau)$ cuts off to zero when $\tau$ is larger than the time $T$ since the start of the birth-death process. The normalizing factor of~\eqref{eq:ptau} is then found to be $(\lambda-\mu\e^{-(\lambda-\mu)T})^2/(\lambda(\lambda-(\lambda-2\mu)\e^{-2(\lambda-\mu)T}-2\mu\e^{-(\lambda-\mu)T}))$.
	Since the normalizing factor converges exponentially towards one as $T$ increases, the asymptotic distribution \new{\eqref{eq:ptau}} is also a valid approximation for populations of finite age $T$ in the regime $(\lambda-\mu)T\gg1$. 
	

	\section{Branch length distribution: interior branches above pendant branches}
	\label{sec:branches_interior_above_pendant}
	
\begin{figure}
		\centering
		A\includegraphics[width=0.3\textwidth]{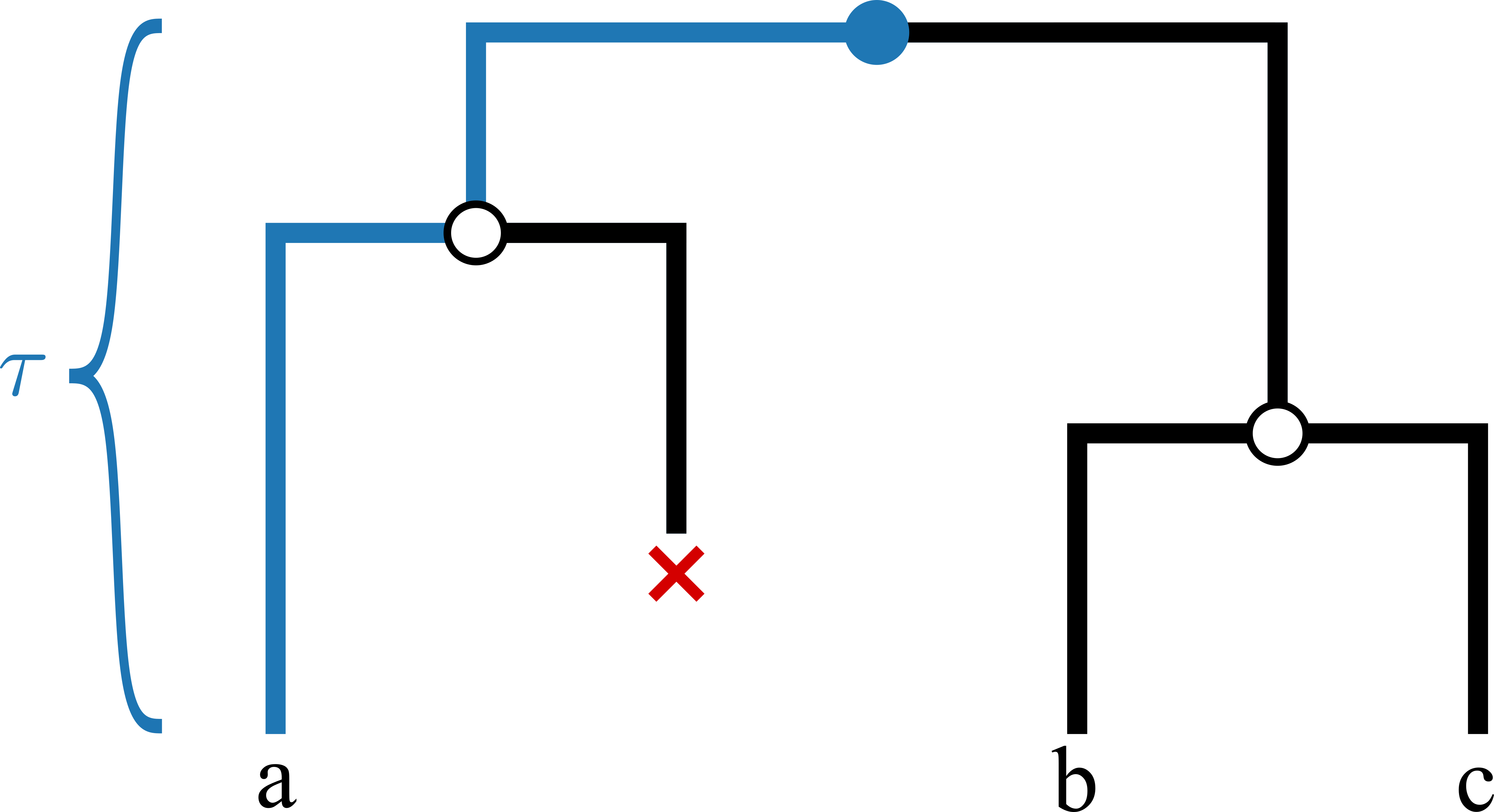}\hspace{.01\textwidth}
		B\includegraphics[width=0.3\textwidth]{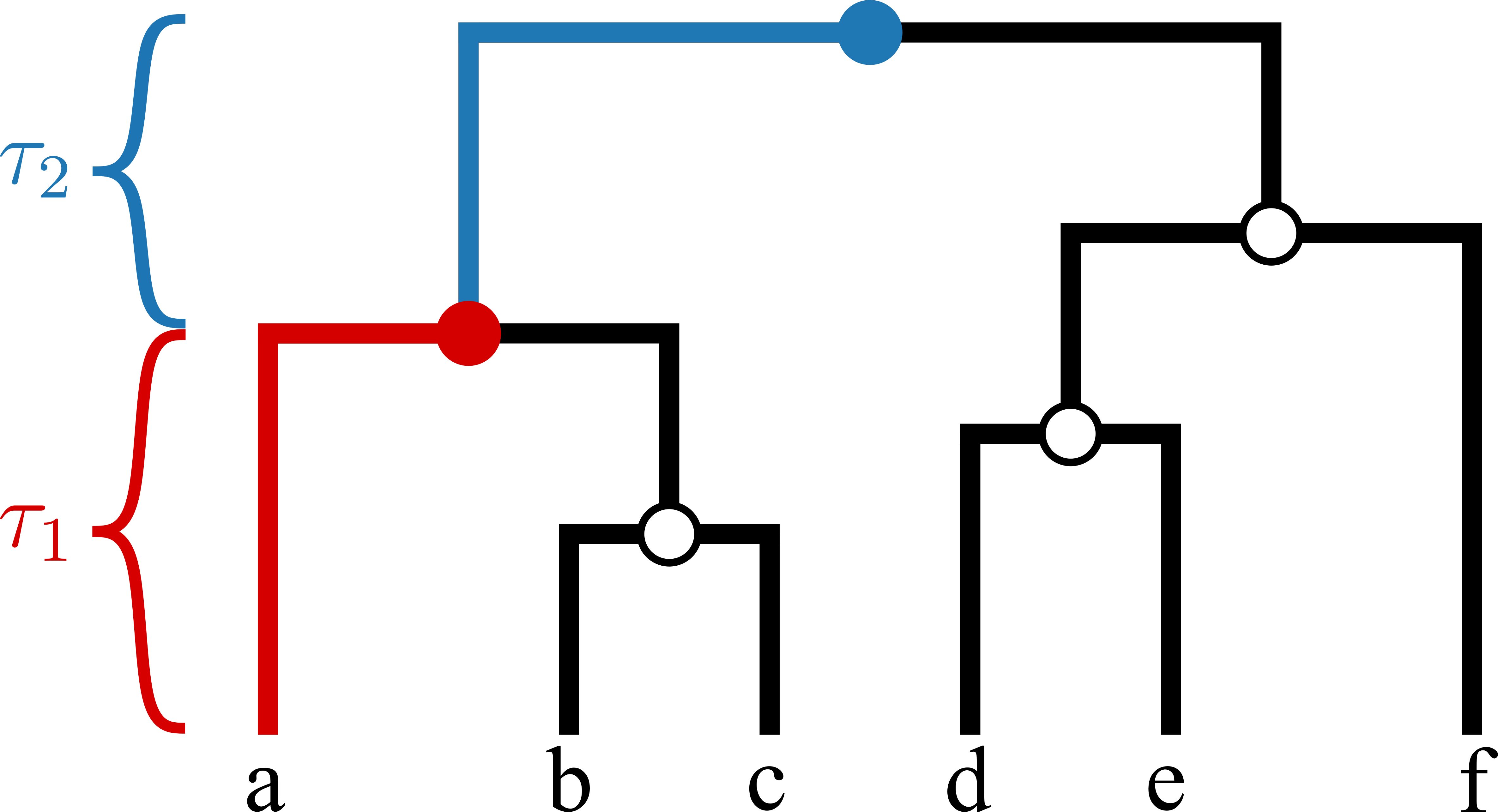}\hspace{.01\textwidth}
		C\includegraphics[width=0.3\textwidth]{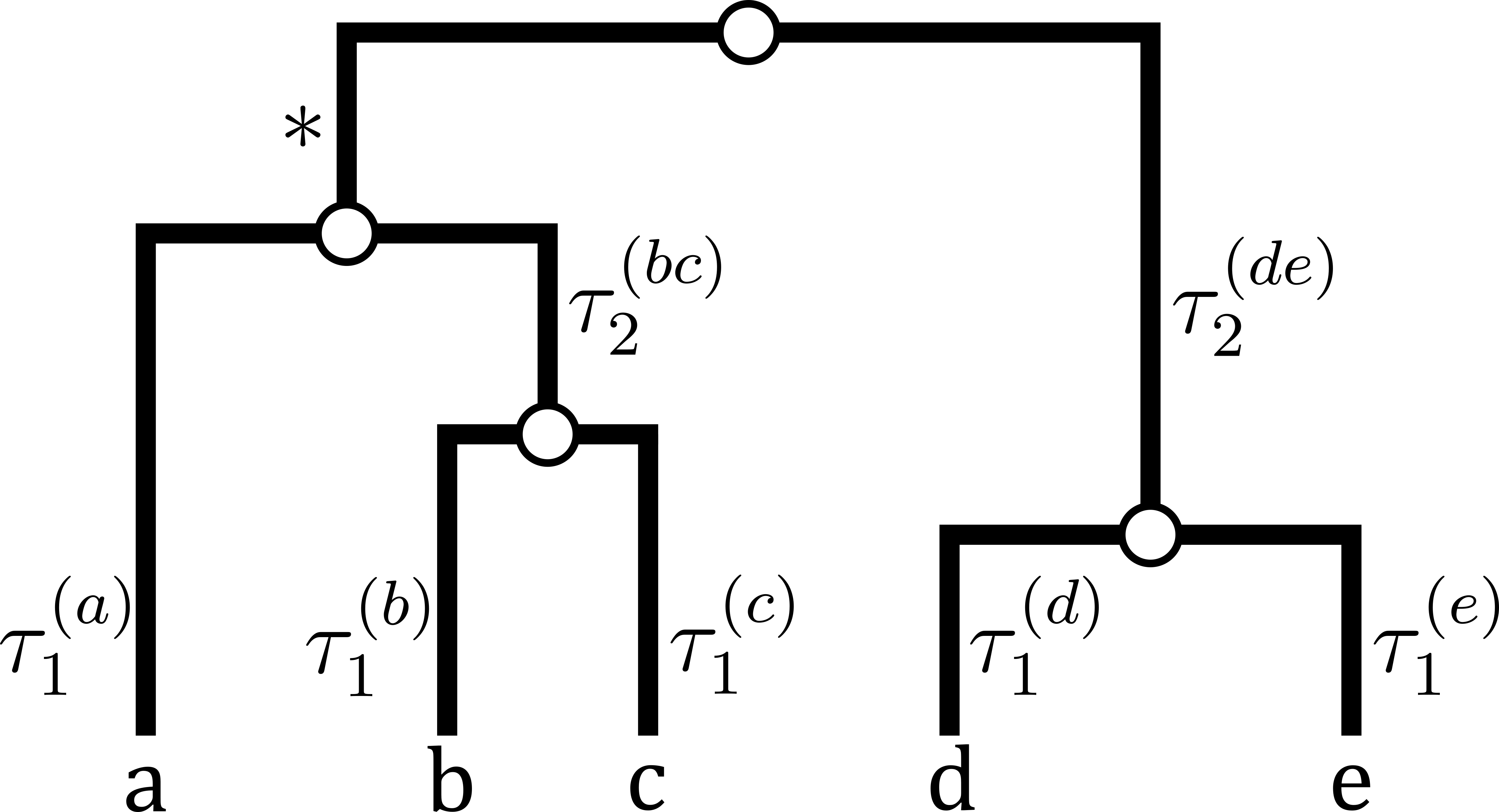} \\
		\vspace{\baselineskip}
		D\includegraphics[width=0.45\textwidth]{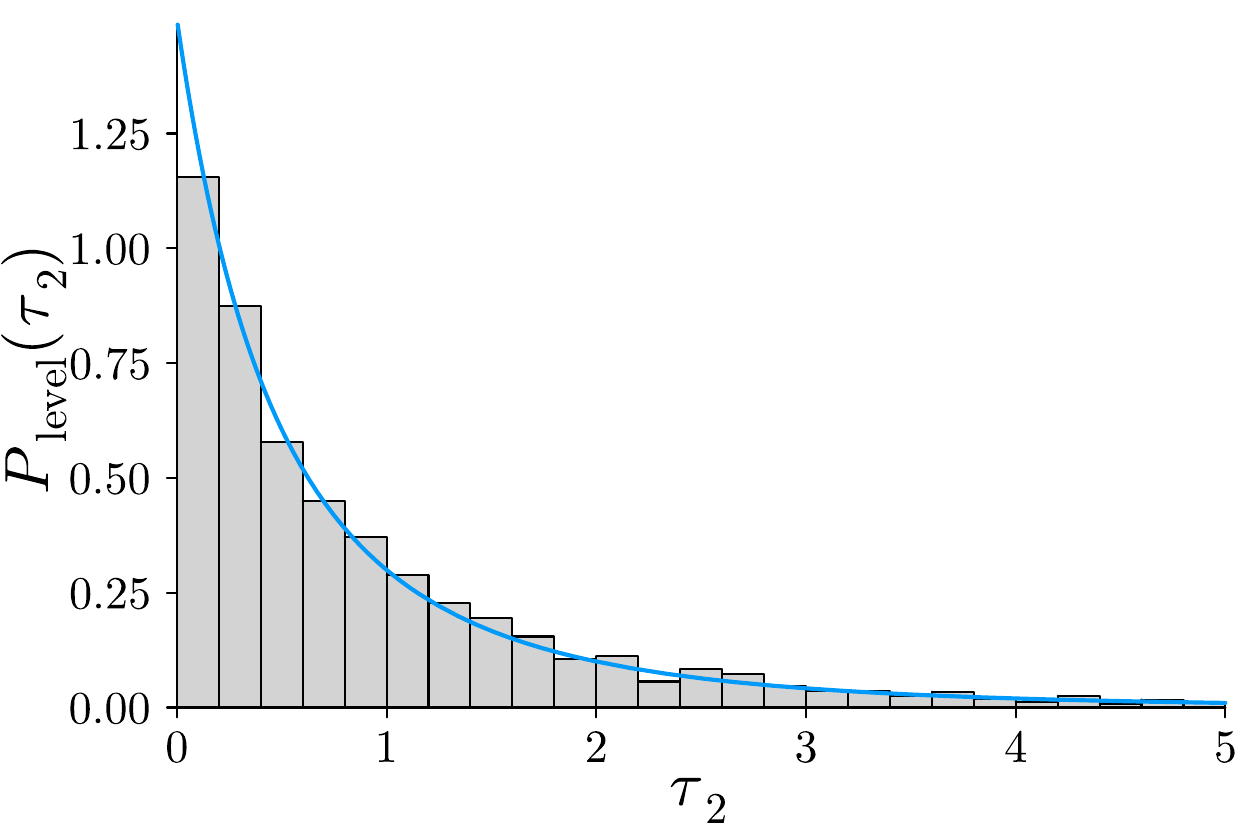}\hspace{.05\textwidth}
		E\includegraphics[width=0.45\textwidth]{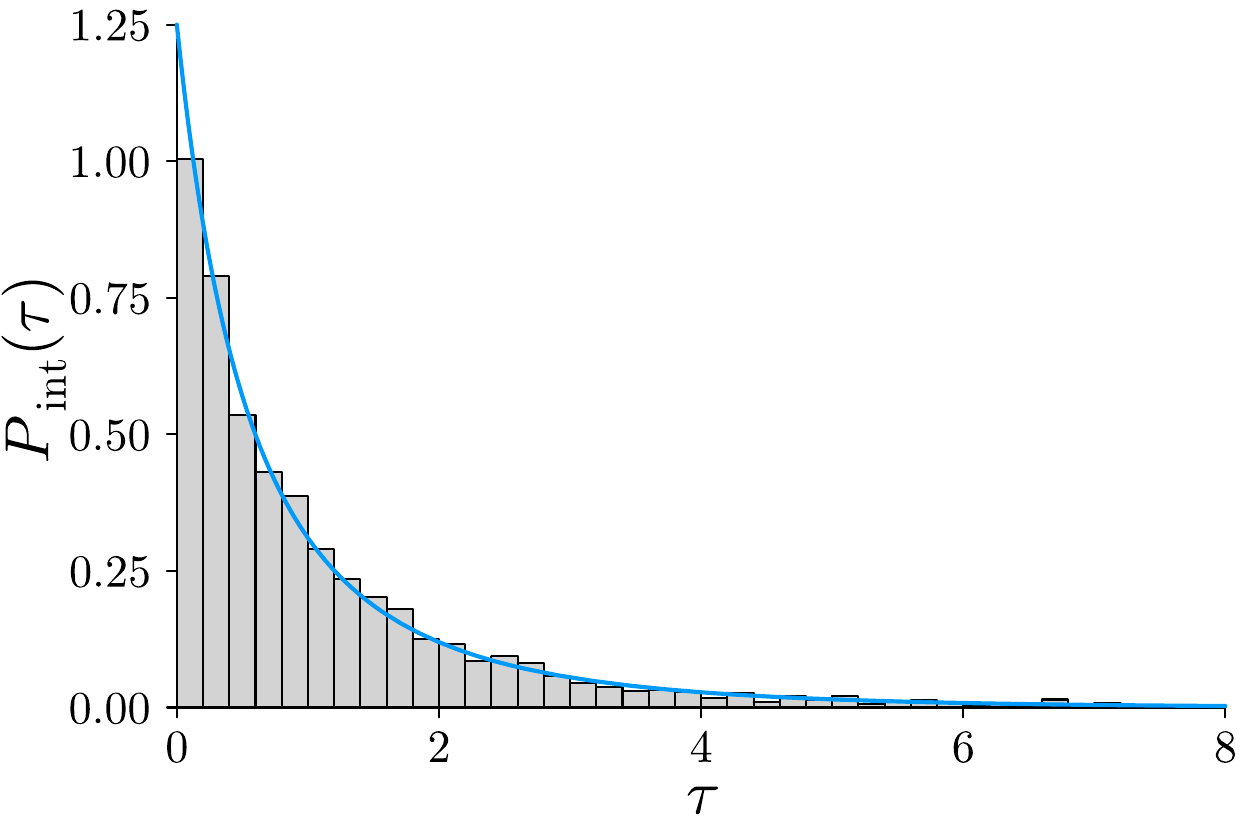}
		\caption{\textbf{Statistics of pendant and interior branches.}
			(A) This schematic phylogenetic tree shows birth events (circles) and death events ($\times$). The pendant branch of length $\tau$ on the left arose from a birth event in the past (solid circle), where one offspring (shown left) has exactly one extant descendant, and the other offspring (shown right) has at least one extant descendant. This particular pendant branch has one generation between its origin at time $\tau$ measured from the time of observation in the past and the present (not counting the birth event at $\tau$). 
			(B) A birth event (top solid node) leading to a level-two branch of length $\tau_2$ and a level-one (pendant) branch of length $\tau_1$. 
			(C) A schematic phylogenetic tree with branches of different levels, see text. 
			(D) The distribution of branch lengths $\tau_2$ of second-level branches in the tree of a simulated population grown from a single individual over time $T=25$ as described in the text (grey bars), compared to the theoretical probability density function~\eqref{eq:ptau1tau2} in the asymptotic case of large $T$ marginalized over the pendant branch lengths $\tau_1$ (blue line).
			(E) The distribution of interior branch lengths in the same tree as before (grey bars), compared to the theoretical density function~\eqref{eq:ptau_interior} (blue line).  
		}\label{fig:interior}
	\end{figure}

	This approach to pendant branches can be extended to the interior branches of a phylogenetic tree. We proceed to go up the phylogenetic tree and first consider branches one level up from pendant branches. We call branches that end in an internal node that is the beginning of a pendant branch \emph{level-two branches}. Correspondingly, pendant branches are \emph{level-one branches}. Figure \ref{fig:interior}B shows the situation: a pair of successive level-two and level-one branches of lengths $\tau_2$ and $\tau_1$, respectively, arose in a birth event at time $\tau_2+\tau_1$ \new{(with time running backwards as described above)}, where one of the two offspring has extant descendants (d-f), and the other has no extant descendants apart from the offspring of individuals arising in another birth event at time $\tau_1$, which gives rise to two offspring. Again, one of these has at least one extant descendant (b-c), and the other has exactly one (a), establishing the pendant branch of length $\tau_1$, see Fig. \ref{fig:interior}B. 
	
	\new{We proceed analogously to pendant branches: A a pair of successive level-two and level-one branches is created by 
	a birth event at time $\tau_1+\tau_2$, the rate of which we know. This rate is multiplied by the probabilities
	that a level-two and level-one branch are created (as specified above).} 
	Since the probability of a birth event at time $\tau_1+\tau_2$ \new{before observation} was 
	$\lambda\e^{-(\lambda-\mu)(\tau_1+\tau_2)}$ (up to a normalizing prefactor, see below, and per small time interval), the relative number of level-two and level-one branches 
	of lengths $\tau_2$ and $\tau_1$ per small time intervals is thus
	\begin{equation}
		\label{eq:ptau1tau2}
		P_\text{level}(\tau_1,\tau_2)
		= 2 \lambda\e^{-(\lambda-\mu)(\tau_1+\tau_2)} p_{1}(\tau_1) \left(1-p_0(\tau_1)\right) 2\lambda p^{(\tau_1)}_{1}(\tau_2) \left(1-p_0(\tau_1+\tau_2)\right)  \ .
	\end{equation}
	$p^{(\tau_1)}_{1}(\tau_2)$ \new{is given by~\eqref{eq:p1_tau1} and} denotes the probability that all birth events taking place along the level-two branch of length $\tau_2$ lead to additional lineages that die out by the time the population is observed. More precisely, it is the probability that a given individual alive at time $\tau_1+\tau_2$ has exactly one descendant alive at time $\tau_1$ that will have one or more extant descendants.
	
	Equation \eqref{eq:ptau1tau2} gives the joint probability density function of branch lengths of pendant and level-two branches up to a normalization factor, which again turns out to be one in the limit of large times $T$. Figure~\ref{fig:interior}D compares this result to numerical simulations. A population with birth rate $\lambda=1$ and death rate $\mu=0.75$ was grown from a single individual until time $T=25$ (where the population contained $4460$ individuals), and the phylogenetic tree was reconstructed. The lengths of level-two branches were collected in a histogram and compared to the joint probability $P_\text{level}(\tau_1,\tau_2)$ given by~\eqref{eq:ptau1tau2} marginalized with respect to $\tau_1$.
	
	This approach can be applied to higher levels in the phylogenetic tree. The joint 
	probability density function of branch lengths $\tau_1,\tau_2$ and $\tau_3$ in a lineage with level-one, -two, and -three branch is 
	\begin{equation}
		\label{eq:ptau1tau2tau3}
		P_\text{level}(\tau_1,\tau_2,\tau_3)
		= 2\lambda \e^{-(\lambda-\mu)(\tau_1+\tau_2+\tau_3)} p_{1}(\tau_1) \left(1-p_0(\tau_1)\right) 2\lambda p^{(\tau_1)}_{1}(\tau_2) \left(1-p_0(\tau_1+\tau_2)\right) 2\lambda p^{(\tau_1+\tau_2)}_{1}(\tau_3) \left(1-p_0(\tau_1+\tau_2+\tau_3)\right) \ .
	\end{equation}
	These densities describe the statistics of branches as one goes up the tree from the different leaves; starting from one particular leaf gives one set of branch lengths $\tau_1,\tau_2,\ldots$, and starting from another leaf gives another set of branch lengths. 
	However, at higher levels than the first, not all branches can be uniquely assigned to a specific level. For instance, the branch marked $\ast$ in Figure~\ref{fig:interior}C contributes both to the distribution of $\tau_2$ (going up the tree from the leaf $\brancha$) and to the distribution of $\tau_3$ (going up from the leaves marked $\branchb$ or $c$).  
	
	\section{Branch length distribution: interior branches}
	\label{sec:branches_interior}
	As a next step, we derive the distribution of internal branch lengths, i.e., all branches that are not pendant \new{(nor the root branch leading from the start of the birth-death process to the first bifurcation)}. To this end, we calculate the joint probability of an interior branch starting at time $\tau_s$ and ending at $\tau_e<\tau_s$. As before, the probability per unit time of a birth event occurring at time $\tau_s$ is proportional to $\lambda\e^{-(1-q)\tau_s}$.
	Focusing for now on only one of the two clades that started at $\tau_s$ (called $A$), the probability of it running until $\tau_e$ with only one extant descendant 
	at the time of observation is $p_1^{(\tau_e)}(\tau_s-\tau_e)$. The probability per small time interval of the branch ending at $\tau_e$ is equal to the probability per small time interval of a birth event at that time where none of the two emerging clades dies out. This is given by $\lambda(1-p_0(\tau_e))^2$.
	
	The second clade (called $B$)  emerging at $\tau_s$ must not die out, which happens with probability $1-p_0(\tau)$. The probability that both branches emerging at $\tau_s$ end at a small interval around $\tau_e$ vanishes quadratically with the time interval and can be neglected. Either clade can be called $A$, which leads to an overall factor of two.

	Putting these elements together, the probability density function of an interior branch running from $\tau_s$ to $\tau_e$ (up to a normalization factor found to be one) is given by
	\begin{align}
			P_\text{int}(\tau_s,\tau_e) &= \lambda\e^{-(\lambda-\mu)\tau_s} p_1^{(\tau_e)}(\tau_s-\tau_e) \lambda\left(1-p_0(\tau_e)\right)^2 2\left( 1 - p_0(\tau_s) \right) \nonumber \\
			&= 2\lambda^2(\lambda-\mu)^3 \frac{\e^{(\lambda-\mu)(\tau_s+\tau_e)}}{(\lambda\e^{(\lambda-\mu)\tau_s}-\mu)^3} \ .  \label{eq:int_branch_propto}
		\end{align}
	
	Marginalizing over $\tau_s$ for $\tau=\tau_s-\tau_e$
	results in the density of interior branch lengths in the entire tree
	\begin{align}
		\label{eq:ptau_interior}
		P_\text{int}(\tau) &= (\lambda-\mu)^2 \frac{2\lambda-\mu\e^{-(\lambda-\mu)\tau}}{\left(\lambda\e^{(\lambda-\mu)\tau}-\mu\right)^2}.
	\end{align}
	Figure~\ref{fig:interior}E compares this result to the empirical distribution of a tree reconstructed from a population of age $T=25$ grown from a single individual with birth rate $\lambda=1$ and death rate $\mu=0.75$ as described above.
	
	So far, we have again only considered the asymptotic case of $T\rightarrow\infty$. For finite $T$, the probability of $\tau_s>T$ is $0$, resulting in a normalization factor of $(\lambda-\mu\e^{-(\lambda-\mu)T})^2/(a(1-\e^{-(\lambda-\mu)T}))^2$ for \eqref{eq:int_branch_propto}. Again, this factor decays exponentially towards one as the population age $T$ increases, justifying the asymptotic results also as an approximation for populations of finite age with $(\lambda-\mu)T\gg1$.

	\section{Finite sampling}
	\label{sec:sampling}
	
	Usually, only a subset of individuals sampled from the population is available to construct the phylogenetic tree from, and this sampling affects the statistics of branch lengths. To model the process of sampling from the extant population we follow the standard approach and choose extant individuals at the time of observation independently and with equal probability $\rho$~\cite{yang_bayesian_1997,stadler_sampling-through-time_2010,stadler_distribution_2012}. The preceding results on branch length distributions can be easily generalized to a sampling probability $\rho<1$ using the probabilities that a clade of a particular age has no (or one)
	member \emph{which is sampled} at a particular time. 
	
	Yang and Rannala~\cite{yang_bayesian_1997} have calculated the probabilities that a birth-death process starting with a single individual at time zero has no \emph{sampled} offspring or a single \emph{sampled} offspring, respectively, as
	\begin{align}
		\label{eq:p0_sampling}
		p_0^{(\rho)}(\tau) &= 1 - \frac{\rho(\lambda-\mu)}{\rho\lambda + \left( (1-\rho)\lambda-\mu\right)\e^{-(\lambda-\mu)\tau} } \\
		\label{eq:p1_sampling}
		p_1^{(\rho)}(\tau) &=\frac{\rho(\lambda-\mu)^2\e^{-(\lambda-\mu)\tau} }{(\rho\lambda + \left( (1-\rho)\lambda-\mu\right)\e^{-(\lambda-\mu)\tau} )^2} \ .
	\end{align}
	These results can be derived by summing over the probabilities of clade sizes $n$ at time $\tau$ (equation~\eqref{eq:pn}) weighted with $(1-\rho)^n$ and \new{$n\rho(1-\rho)^{n-1}$}, respectively.
	
	\new{Based on these probabilities, we can repeat the derivations from Section~\ref{sec:birth_death} to compute the probability $p_1^{(\rho)(\tau_e)}(\tau)$ that a clade starting with a single individual $\tau_e+\tau$ before the time of observation has one descendant alive at time $\tau$ (with an open future fate) and none of the additional lineages arising between $\tau_e+\tau$ and $\tau$ leave are \textit{sampled} at time $0$.} Replacing $p_0(\tau)$ with $\eqref{eq:p0_sampling}$ in master equation~\eqref{eq:master_p1} yields 
	\begin{align}
		\label{eq:p1_tau_e_sampling}
		p_1^{(\rho)(\tau_e)}(\tau) &= \frac{\left(\rho\lambda - (\mu-(1-\rho)\lambda)\e^{-(\lambda-\mu)\tau_e}\right)^2\e^{-(\lambda-\mu)\tau}}{\left(\rho\lambda - (\mu-(1-\rho)\lambda)\e^{-(\lambda-\mu)(\tau_e+\tau)}\right)^2} \ .
	\end{align}
	Note that in this case, $p_1^{(\rho)}(\tau)=\rho p_1^{(\rho)(\tau_e=0)}(\tau)$, because we need to include the probability that the lineage running until $\tau_e=0$ is actually sampled.
	
	To derive the probability density function of branch lengths under sampling, these probabilities can be used directly in the densities of pendant and interior branch lengths at full sampling ($\rho=1$), \eqref{eq:ptau}-\eqref{eq:int_branch_propto}, which depend only on $p_0$, $p_1$, and the relative population sizes at different times. \new{Replacing $p_0(\tau)$, $p_1(\tau)$ and $p_1^{(\rho)}(\tau)$ with their respective counterparts \eqref{eq:p0_sampling}-\eqref{eq:p1_tau_e_sampling} thus fully accounts for finite sampling, as the population size in the past is not affected by sampling of the extant population. In this way}, we obtain
	\begin{align}
		\label{eq:p_pend_f}
		P^{(\rho)}_\text{pend}(\tau) &= 2\lambda\rho(\lambda-\mu)^3 \frac{\e^{(\lambda-\mu)\tau}}{\left(\lambda \rho\e^{(\lambda-\mu)\tau}+\lambda(1-\rho)-\mu\right)^3 } \\
		\label{eq:p_int_joint_f}
		P_\text{int}^{(\rho)}(\tau_s,\tau_e)
		&= 2(\lambda\rho)^2(\lambda-\mu)^3 \frac{\e^{(\lambda-\mu)(\tau_s+\tau_e)}}{\left(\lambda \rho\e^{(\lambda-\mu)\tau_s} + \lambda(1-\rho)-\mu\right)^3} \\
		\label{eq:p_int_f}
		P_\text{int}^{(\rho)}(\tau)
		&= (\lambda-\mu)^2 \frac{2\lambda \rho - \e^{-(\lambda-\mu)\tau}\left(\mu-\lambda(1-\rho)\right)}{\left(\lambda(1-\rho)-\mu+\lambda \rho\e^{(\lambda-\mu)\tau}\right)^2} \ .
	\end{align}
	A normalizing factor equalling $\rho^{-1}$ in the limit of large $T$ applies to all three densities and is included in the expressions above. The result for pendant branches coincides with 
the result of Stadler and Steel~\cite{stadler_distribution_2012} at a fixed number of samples. 
However, their result was restricted to a regime of high death rate or high sampling probability $1\geq \mu/\lambda \geq 1-\rho$. The density function \eqref{eq:p_pend_f} also holds outside these limits.
	
\begin{figure}[tbh]
	\centering
	A\includegraphics[width=0.6\textwidth]{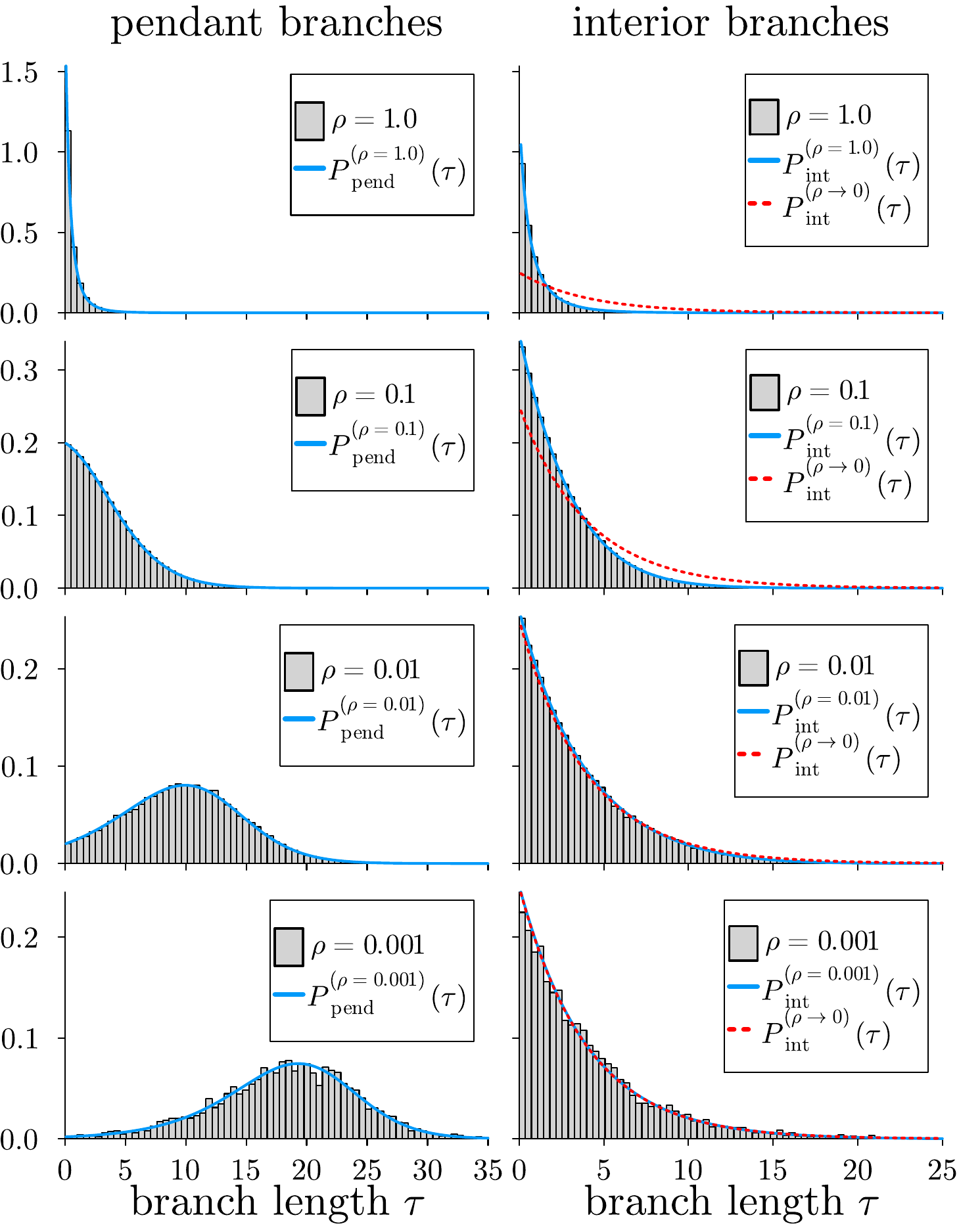}
	B\;\includegraphics[width=0.33\textwidth]{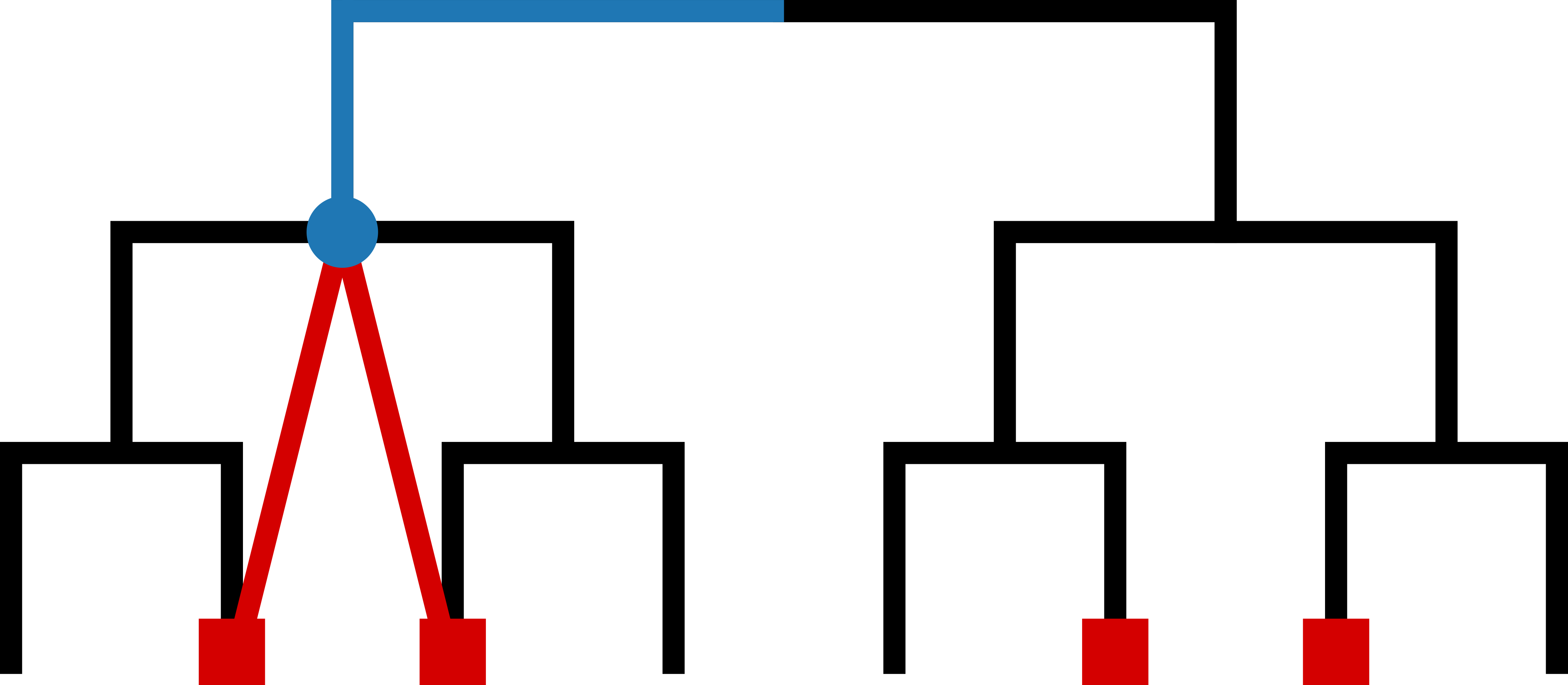}
	\caption{\textbf{Pendant and interior branches at different sampling probabilities $\rho$.}
		(A) We compare the analytical results for the probability density of the length of pendant and interior branches under sampling (\eqref{eq:p_pend_f} and \eqref{eq:p_int_f} respectively, solid blue lines) to numerical simulations. 
		For the latter, we show the distribution of branches from 100 trees. For every tree, we simulated the growth of a population with $\lambda=1$ and $\mu=0.75$ over time $T=40$, restarting when the population died out or the final population size was not within $2\times 10^4$ and $3 \times 10^5$ (grey bars). The first row of plots shows the branch length distribution of the full trees. For the next row of plots, individuals were sampled with probability $\rho=0.1$ and the corresponding trees were reconstructed. For the third row, each of the previous samples was sampled again with probability $0.1$, corresponding to an overall sampling probability of $\rho=0.01$. The same was repeated for the fourth row of plots, yielding $\rho=0.001$. Every tree contributes equally to the empirical distributions depicted.  The first column shows the distributions of pendant branch lengths, which shift to longer branches as $\rho$ decreases. The second column of plots shows the distribution of interior branches, quickly reaching the asymptotic distribution~\eqref{eq:p_int_f0} (red dashed line) as $\rho$ becomes small.
		(B) A toy example of how sampling 
		affects the length of pendant branches, see text. 
	}\label{fig:sampling}
\end{figure}
	
	The lengths of pendant and interior branches behave very differently under sampling. Pendant and interior branch lengths are compared in Figure \ref{fig:sampling}A. The rows correspond to different sampling probabilities $\rho$, the first column shows pendant, the second column interior branches. Histograms show the branch length distributions of $100$ simulations of trees, while the blue lines depict the analytic results. Pendant branches in the first column get longer with decreasing sampling probability $\rho$. On the other hand, interior branches quickly reach an asymptotic distribution as $\rho$ is decreased and then do not get any longer as $\rho$ is decreased further, see the second column of Figure \ref{fig:sampling}A. The red dashed line depicts the asymptotic state, which can be derived as
	\begin{align}
		\label{eq:p_int_f0}
		P^{(\rho\rightarrow 0)}_\text{int}(\tau) = \lim_{\rho\rightarrow 0}P^{(\rho)}_\text{int}(\tau) = (\lambda-\mu) \e^{-(\lambda-\mu)\tau}
	\end{align}
	and is thus an exponential distribution with parameter $\lambda-\mu$. 
	
	The difference between pendant and interior branches under sampling has a simple qualitative explanation, which we explore within a toy problem. Consider a reconstructed tree which is perfectly balanced; both clades defined by each bifurcation have the same size. An example is shown in Fig. \ref{fig:sampling}B. Then a fraction $\rho$ of the extant nodes is sampled. Under sampling, a pair of pendant branches linked to a shared ancestor is defined by the birth of a clade (shown in blue in Fig. \ref{fig:sampling}B) whose size $n$ at the time of observation is defined by $n\rho \approx 2$ (so two of its members are expected to be sampled). The number of branch levels between this founding 
	birth event and the time of observation is $\ln_2(n)$, so the pendant branch lengths are increased under sampling by a factor of $1-\ln_2 \rho\approx -\ln_2 \rho$ relative to the fully sampled case. (This argument assumes that on the fully sampled reconstructed tree, pendant and interior branches are approximately of the same length.) On the other hand, branches that are interior in the reconstructed tree (i.e. lie above birth event shown in blue in Fig. \ref{fig:sampling}B) have their lengths unchanged by sampling. The distribution of interior branches only changes because under sampling, fewer and older branches are counted as interior. Older branches tend to be a little longer (see equation~\eqref{eq:int_branch_propto}), because any additional lineages that arose along them had more time to die out until the time of observation. 
	
	To show this quantitatively, we compute the cumulative distribution function of the end $\tau_e$ of interior branches
	\begin{align}
		\int_0^{\tau_e}\d\tau'_e\int_{\tau'_e}^\infty\d\tau_s P_\text{int}^{(\rho)}(\tau_s,\tau'_e) &= \lambda\rho \frac{1-\e^{-(\lambda-\mu)\tau_e}}{\lambda\rho + \left(\lambda(1-\rho)-\mu\right)\e^{-(\lambda-\mu)\tau_e}} \ .
		\label{eq:cum_dist_tau_e}
	\end{align}
	For any finite $\tau_e$, this distribution approaches zero with decreasing sampling probability $\rho$. This means that the ends of interior branches lie asymptotically far away from the time of observation when $\rho$ approaches zero. Setting equation \eqref{eq:cum_dist_tau_e} equal to $1/2$ and solving for $\tau_e$ yields the median interior branch end. In the regime $\e^{-(\lambda-\mu)\tau_e}\ll1$, the median is given by $\ln\left( (\lambda-\mu)/\rho\right)/(\lambda-\mu)$, reproducing the logarithmic dependence on the sampling probability found in the toy model. 
	
	Stadler \cite{stadler_incomplete_2009} found that the probability distribution of bifurcation times is not unique to a particular set of parameters, but that a birth-death process with birth rate $\lambda$, death rate $\mu$ and sampling probability $\rho$ yields the same distribution as a process with parameters $\rho'$, $\lambda'=\rho\lambda/\rho'$, $\mu'=\mu-\lambda(1-\rho/\rho')$. We find the same for the distributions of branch lengths \eqref{eq:p_pend_f}-\eqref{eq:p_int_f}, as $\lambda'-\mu'=\lambda-\mu$, $\rho'\lambda'=\rho\lambda$ and $\lambda'(1-\rho')-\mu'=\lambda(1-\rho)-\mu$ (which is the same argument used in \cite{stadler_incomplete_2009,stadler_how_2013}). Therefore, inference of both growth parameters and the sampling probability from the distribution of branch lengths is impossible. However, an upper bound on the sampling probability can be determined: Given a branch distribution generated with (unknown) parameters $(\lambda, \mu, \rho)$, the parameter space of all sets of $(\lambda',\mu',\rho')$ which lead to the same distribution is restricted to $\mu'\geq0$ and $0<\rho\leq 1$. With the transformation rules above, it is easy to see that these inequalities are only fulfilled if $0<\rho'\leq\min(\rho\lambda/(\lambda-\mu), 1)$. This means that for branch distributions generated with birth rate $\lambda$, death rate $\mu$ and sampling probability $\rho < 1-\mu/\lambda$, there exists a maximal sampling probability $\rho'<1$ for the family of possible parameter sets. This agrees qualitatively with the behavior of pendant branches in Figure \ref{fig:sampling}A: At full sampling, the maximum of the pendant branches is always found at $\tau=0$. With decreasing sampling probability the time $\tau$ of the maximum eventually becomes non-zero and continues to increase monotonically when the sampling probability is decreased further. For a tree with a pendant branch distribution with a maximum at $\tau>0$, no set of parameters $\lambda,\mu$ can be found that together with $\rho=1$ would lead to the observed distribution.

	\section{Branch lengths conditioned on the population size at the time of observation}
	\label{sec:scenario2}
	
	We now turn to the second statistical scenario, where the birth-death process is conditioned on the number $N$ of extant individuals rather than a particular time of observation (scenario ii) above). This case is harder to treat than scenario i), because with a constraint on the final number of individuals, the dynamics of different clades are in general correlated. We set up a framework to calculate branch length distributions based on tracking the dynamics of the birth-death process first backwards in time from the time of observation to the emergence of a particular branch and then again forward in time to the time of observation.
	For pendant branches, this approach recovers the results of Stadler, Steel and collaborators~\cite{stadler_distribution_2012,mooers_branch_2012}, but it also allows treating interior branches. 
	
	We \new{start by illustrating the method with a toy problem: We calculate the distribution of the length of pendant branches when the number of individuals at the time of observation is $N=2$.  
	A key tool is the probability that a birth-death process with birth rate $\lambda$ and death rate $\mu$, starting with $n$ individuals has $N$ individuals \new{after evolving for time $t$}, denoted $p_{(N,n)}^{(\lambda,\mu)}(t)$.
	
	\begin{figure}
		\centering
		\includegraphics[width=0.60\textwidth]{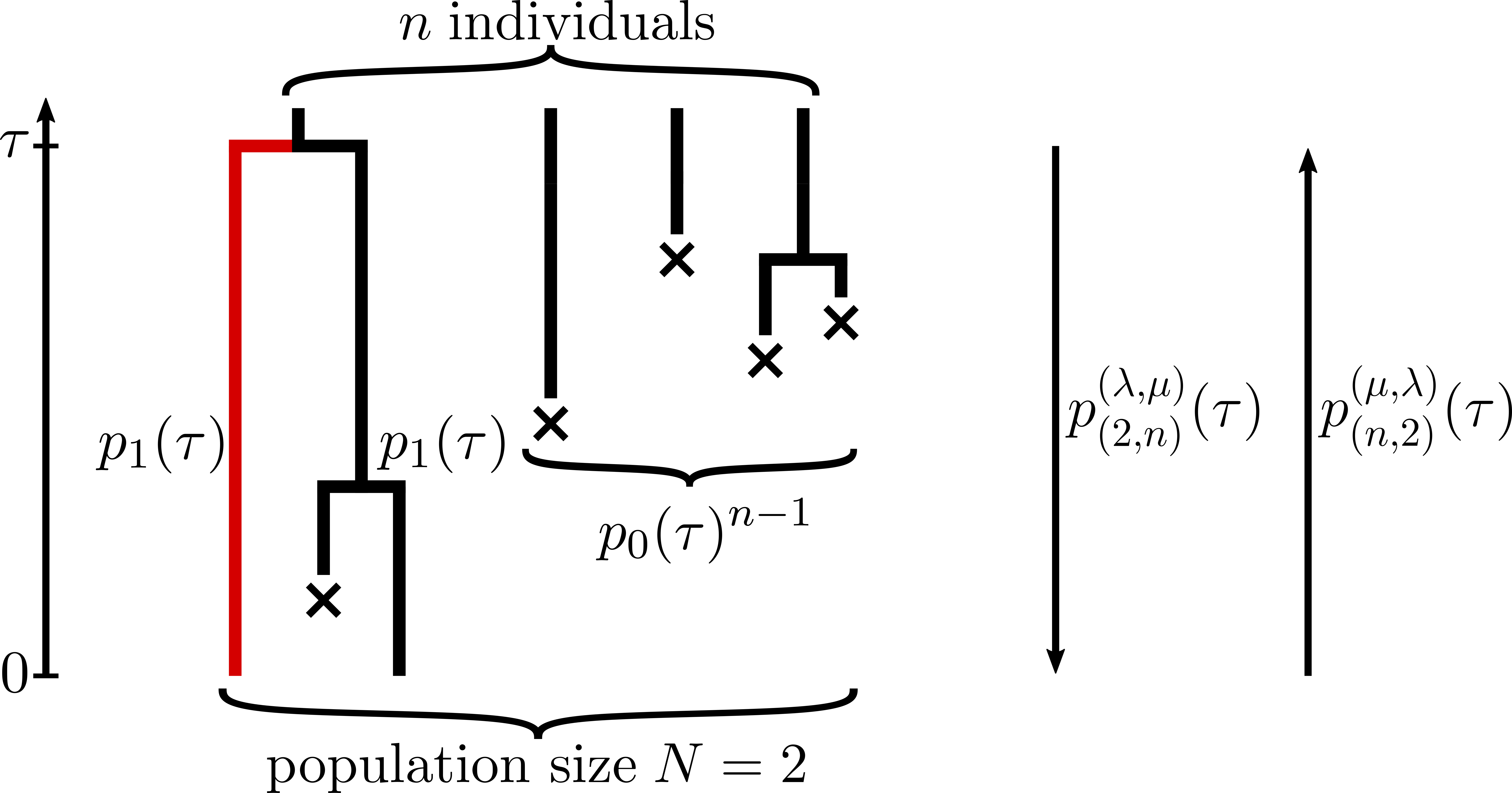}
		\caption{\textbf{Pendant branch formation for populations of size $N=2$:} The red branch marks the pendant branch arising from a birth event at time $\tau$ in the past. All other $n-1$ individuals that were alive at time $\tau$ need to die out before the population is observed, while the second descendant from the birth event needs to survive with clade size one. The arrows on the right illustrate the conditional probabilities of having $N$ extant descendants given that $n$ individuals at time $\tau$ before observation (forwards process) and the probability that given $N$ observed individuals there were $n$ individuals at time $\tau$ before observation (backwards process).
		}\label{fig:propagation_N2}
	\end{figure}
	
	We start by considering the population at the time of observation (with fixed population size $N$) and ask about its population size 
	at a previous time $\tau$. Conditioned on the number of individuals at the time of observation being $N$, the statistics of the number of individuals before the time of observation has been characterized by Ignatieva, Hein and Jenkins~\cite{ignatieva_characterisation_2020}. It 
	is described by that of a birth-death process with time running backwards and with the rates of birth and death exchanged, so the rate of birth is $\mu$ and the rate of death is $\lambda$~\cite{ignatieva_characterisation_2020}. The probability of having $n$ individuals at time $\tau$ before the time of observation is thus $p_{(n,N)}^{(\mu,\lambda)}(\tau)$. This number of individuals $n$ will eventually be summed over.
	
	With $n$ individuals at time $\tau$, the expected number of birth events at time $\tau$ (per small time interval) leading to pendant branches, conditioned on $N=2$ individuals at the time of observation is 
	\begin{align}
		\lambda n p_1^2(\tau) p_0^{n-1}(\tau) \new{\frac{1}{p_{(2,n)}^{(\lambda,\mu)}(\tau)}}.
	\end{align}
	$\lambda n$ gives the probability per small time interval that one of the $n$ individuals divides during this small interval, $p_1^2(\tau)$ is the probability that the two resulting offspring each have one extant offspring at the time of observation, leading to pendant branches of length $\tau$. $p_0^{n-1}(\tau)$ is the probability that all other $n-1$ individuals die out by the time of observation, leading to a population of size $N=2$. The denominator arises from the conditioning on $N=2$ individuals. What we have done is first asking about the statistics of the number of individuals at a certain time $\tau$ in the past, given $N=2$ individuals at some time of observation, and then asking about the probability that a tree with a particular feature (here a pendant branch that emerged at time $\tau$) arises from those forward-time trajectories ending in $N=2$ individuals at the time of observation. We thus first went "backwards in time" from the time of observation to a time $\tau$ before that, and then "forwards in time" to the time of observation. For an illustration, see Figure~\ref{fig:propagation_N2}. 
	
	Up to a normalizing constant, the probability density function of the pendant branch lengths is thus 
	\begin{align}
		\label{eq:toyfinal}
		\sum_{n=1}^{\infty} \lambda n p_1^2(\tau) p_0^{n-1}(\tau) \frac{p_{(n,2)}^{(\mu,\lambda)}(\tau)}{ p_{(2,n)}^{(\lambda,\mu)}(\tau) } \new{= 2 \lambda p_1(\tau)^2 \sum_{n=1}^{\infty} p_0(\tau)^{n-1}=2 \lambda \frac{p_1(\tau)^2}{1-p_0(\tau)}} \ .
	\end{align}
	Here, we have used that $\frac{p_{(n,2)}^{(\mu,\lambda)}(\tau)}{ p_{(2,n)}^{(\lambda,\mu)}(\tau) } = \frac{N}{n}$, which we show in Appendix~\ref{app:cond_N_pendant} (equation \eqref{eq:relative_propagators}). The normalizing constant turns out to be one.}

	We find the same result for a final population size $N\geq2$, see Appendix~\ref{app:cond_N_pendant}.  
	This reproduces the result by Stadler and collaborators on the distribution of pendant branches conditioned on the size $N$ of the population at the time of observation. Curiously, this distribution also coincides with the distribution of pendant branch lengths under scenario i) in the limit of large $T$ (equation~\eqref{eq:ptau}).  
	
	The same approach can also be used to calculate the distribution of interior branch lengths conditioned on the population size. Again we use the birth-death process with time first running backwards to compute the probability that at a time $\tau_s$ in the past there were $n$ individuals, one of which 
	divided establishing a new branch. Then running time forwards again while conditioning on the final population size $N$, we calculate the probability that 
	the new branch ends at $\tau_e$ in another birth event, with both descendants having extant offspring. The conditions that lead to a branch of a particular length are the same manner as in scenario i), the difference is the conditioning on the final population size in two directions of time. The calculation is given 
	in Appendix~\ref{app:cond_N_interior}. 
	
	The resulting distribution of interior branch lengths under scenario ii) depends on $N$. However, the dependency quickly vanishes with $\left(\frac{\lambda}{\mu}p_0(\tau_s)\right)^{N-1}$ when increasing the population size. The asymptotic distribution for large $N$ coincides with the interior branch length distribution under scenario i) in the limit of large times $T$, see Appendix~\ref{app:cond_N_interior}. 
	
	\section{Discussion}
	\label{sec:discussion}
	
	We have calculated the statistics of branch lengths (in calendar time) of phylogenetic trees generated by a constant-rate birth-death process and reconstructed from the population at a single moment in time. 
	We look at two distinct statistical scenarios: sampling at a given moment in time since the birth-death process started and sampling when to population has a given size $N$. A particular focus is on how branch lengths are affected when the population is sampled at different sampling probabilities and only a random subset of individuals can be used to reconstruct the tree. 
	
	For the first scenario, we generalize the approach of Paradis~\cite{paradis_distribution_2016} of observing the population at a given time to a finite sampling probability. We find that pendant branches and interior branches behave very differently under sampling: only pendant branch lengths increase indefinitely as the sampling probability is decreased to zero, whereas the distribution of interior branches quickly reaches an asymptotic distribution. \new{An intuitive explanation of this effect is given in Section~\ref{sec:sampling}}. This behaviour allows placing bounds on the model parameters (birth and death rates, sampling probability) on the basis of empirical trees. An interesting consequence of sampling at a low probability concerns the number of generations along interior branches (birth events which have not led to additional sampled clades): even at low sampling probabilities, the number of generations along interior branches can be small due to the short branch lengths. \new{In contrast to macroevolution, where mutations arise at some rate per unit time, under cellular reproduction mutations can also occur at discrete moments in time at birth events~\cite{spisak2023disentangling}.  
	This means that the number of mutations along a branch can be shaped by the (few) birth events along that branch, which affects the statistics of mutations and leads to a compound Poisson statistics replacing the Poisson statistics that typically arises from the continuous accumulation of mutations. The interplay between reproduction and mutations also affects the statistics of mutations along a lineage~\cite{cheek2023ancestral}. In~\cite{dieselhorst_phylodynamic_2024}, we use the mutation statistics along branches to infer the parameters of the underlying birth-death model.}

	For the second scenario, a fixed given final population size $N$, we develop a statistical framework based on first going backwards in time from a fixed population size, and then forwards again while conditioning on that population size. 
	The latter step entails calculating the probability of a population with a given final size \emph{and} a particular feature of its phylogenetic tree (such as a branch of a given length), relative to the probability of a population with a certain size. 
	The combination of first going backwards and then forwards in time is similar in spirit to the celebrated Keldysh--–Schwinger formalism in non-equilibrium quantum physics~\cite{kamenev_field_2011}. The ``going backwards in time"-aspect of this approach is shared with coalescent approaches, like the Kingman-coalescent~\cite{kingman1982genealogy}. However, the statistics we compute follows that of the original birth-death process which generated the tree. It thus includes 
	the effects of lineages that died out, and the distinct behaviour of pendant and interior branches under sampling, which are generally missing in coalescent approaches. 

	Our approach simplifies the derivation of the distribution of pendant branch lengths by Stadler, Steel and collaborators~\cite{stadler_distribution_2012,mooers_branch_2012} and allows also to treat the length distribution of interior branches. The interior branch distributions of the asymptotic cases of both scenarios (large times for scenario i) and large population sizes for ii)) turn out to coincide. In case of pendant branches, the asymptotic distribution in the limit of large times since the start of the birth-death process coincides with the distribution in trees conditioned on population size $N$, even when $N$ is not large. 
	
	\new{These results all hinge upon the assumption of a constant-rate birth-death model. Changes in birth and death rates can occur over time or between different clades, and this would affect the statistics of branch lengths. In turn, such changes in the statistics of branch lengths allow one in principle to infer rate changes from data~\cite{dieselhorst_phylodynamic_2024}. Also, in a cellular setting, the rates of birth depend on the cell cycle~\cite{mulberry2025bayesian}. Calculating the resulting branch length distributions remains a task to be undertaken in future work.} 
	
	\begin{acknowledgments}
		This work was funded by the Deutsche Forschungsgemeinschaft (DFG, German Research Foundation) grant SFB1310/2 - 325931972. Many thanks to Tibor Antal, Samuel Johnston, Joachim Krug, Thomas Wiehe for discussions, and to Tanja Stadler for comments throughout and especially for helpful advice on section~\ref{sec:sampling}. We thank the computing center of the University of Cologne RRZK for computing time on the DFG-funded CHEOPS cluster (grant number INST 216/512/1FUGG). 
	\end{acknowledgments}

	\bibliographystyle{abbrvnat}
	\bibliography{birthdeath24}
	
	\appendix
	
	\section*{Appendix: Branch length distributions conditioned on the population size at the time of observation}
	\label{app:cond_N}
	
	The probability of a birth-death process with birth rate $\lambda$ and death rate $\mu$ taking a population of size $n>0$ to size $N\geq0$ over time $t$ is given by~(equation 8.47 in \cite{bailey_elements_1991})
	\begin{align}
		\new{p_{(N,n)}^{(\lambda,\mu)}(t)} &= \sum_{j=0}^{\min(n,N)} {{n}\choose{j}} {{n+N-j-1}\choose{n-1}} \new{ p_0^{n-j} \left(\frac{\lambda}{\mu}p_0\right)^{N-j} (1-p_0-\frac{\lambda}{\mu}p_0)^j }
		\label{eq:P_N_given_n}
	\end{align}
	\new{with shorthand $p_0=p_0(t)=p_0(t,\lambda,\mu)$ (equation~\eqref{eq:p0}). (The explicit dependence of $p_0$ on $\lambda$ and $\mu$ will become relevant later on.)}
	\new{The probability~\eqref{eq:P_N_given_n}} can be derived by considering $n$ clades starting out with a single individual at a particular time and evolving according to \eqref{eq:pn}, and computing the probability that the sum of all clade sizes a time $t$ later is $N$. For $N=0$ equation \eqref{eq:P_N_given_n} simplifies to \new{$p_{(0,n)}^{(\lambda,\mu)}(t)=p_0(\tau)^n$}.
	
	The conditional probability~\eqref{eq:P_N_given_n} can also be used to describe the statistics of the population size $n$ some time $\tau$ in the past, conditioned on a population size $N$ at the time of observation. Ignatieva, Hein, and Jenkins~\cite{ignatieva_characterisation_2020} show that the statistics of $n$ is described by a birth-death process running for time $\tau$, starting with $N$ and with the rates of birth and death exchanged. This result uses a flat prior on the (unknown) age $T$ of the birth-death process. The same flat prior is also used explicitly in the derivation of the pendant branch length distribution conditioned on the final population size by Stadler and collaborators~\cite{,mooers_branch_2012,stadler_distribution_2012}.
	
	An important observation that simplifies the subsequent calculations is that the ratio of conditional probabilities going forwards \new{(from $n$ to $N$ individuals with birth rate $\lambda$ and death rate $\mu$)} and backwards \new{(from $N$ to $n$ individuals with birth rate $\mu$ and death rate $\lambda$)} in time simplifies to 
	\begin{equation}
		\label{eq:relative_propagators}
		\new{\frac{p_{(n,N)}^{(\mu,\lambda)}(\tau)}{ p_{(N,n)}^{(\lambda,\mu)}(\tau)}} = N/n \ .
	\end{equation}
	for all times and rates. This can be derived using the symmetry in the exchange of birth and death rates \new{$p_0(t,\mu,\lambda)=\frac{\lambda}{\mu}p_0(t,\lambda,\mu)$}. Ignoring the binomial coefficients in equation \eqref{eq:P_N_given_n}, the remaining terms are therefore invariant to swapping $\lambda\leftrightarrow\mu$ and $N\leftrightarrow n$. For the binomial coefficients we get ${{N}\choose{j}} {{N+n-j-1}\choose{N-1}}=\frac{N}{n}  {{n}\choose{j}} {{n+N-j-1}\choose{n-1}}$.
	Since the upper limit of the sum in equation \eqref{eq:P_N_given_n} is invariant under exchange of $n$ and $N$, we get the result \eqref{eq:relative_propagators}.
	
	\subsection{Pendant branches}
	\label{app:cond_N_pendant}
	We \new{start by solving} the toy problem of Section~\ref{sec:scenario2} (pendant branch length distribution conditioned on a final population of size $N=2$) in \new{two alternative} ways of increasing generalizability. The \new{second} one generalizes directly to the result for any value of the final population size $N$ and to the goal of this appendix, the distribution of interior branch lengths conditioned on population size. \new{In the following, we will use the shorthand $P(N|n) \equiv \new{p_{(N,n)}^{(\lambda,\mu)}(\tau)}$ and do not denote all dependencies on $\tau$, i.e.\ $p_0=p_0(\tau)$ and $p_1=p_1(\tau)$.}
	
\begin{figure}
		\centering
		A\includegraphics[width=0.3\textwidth]{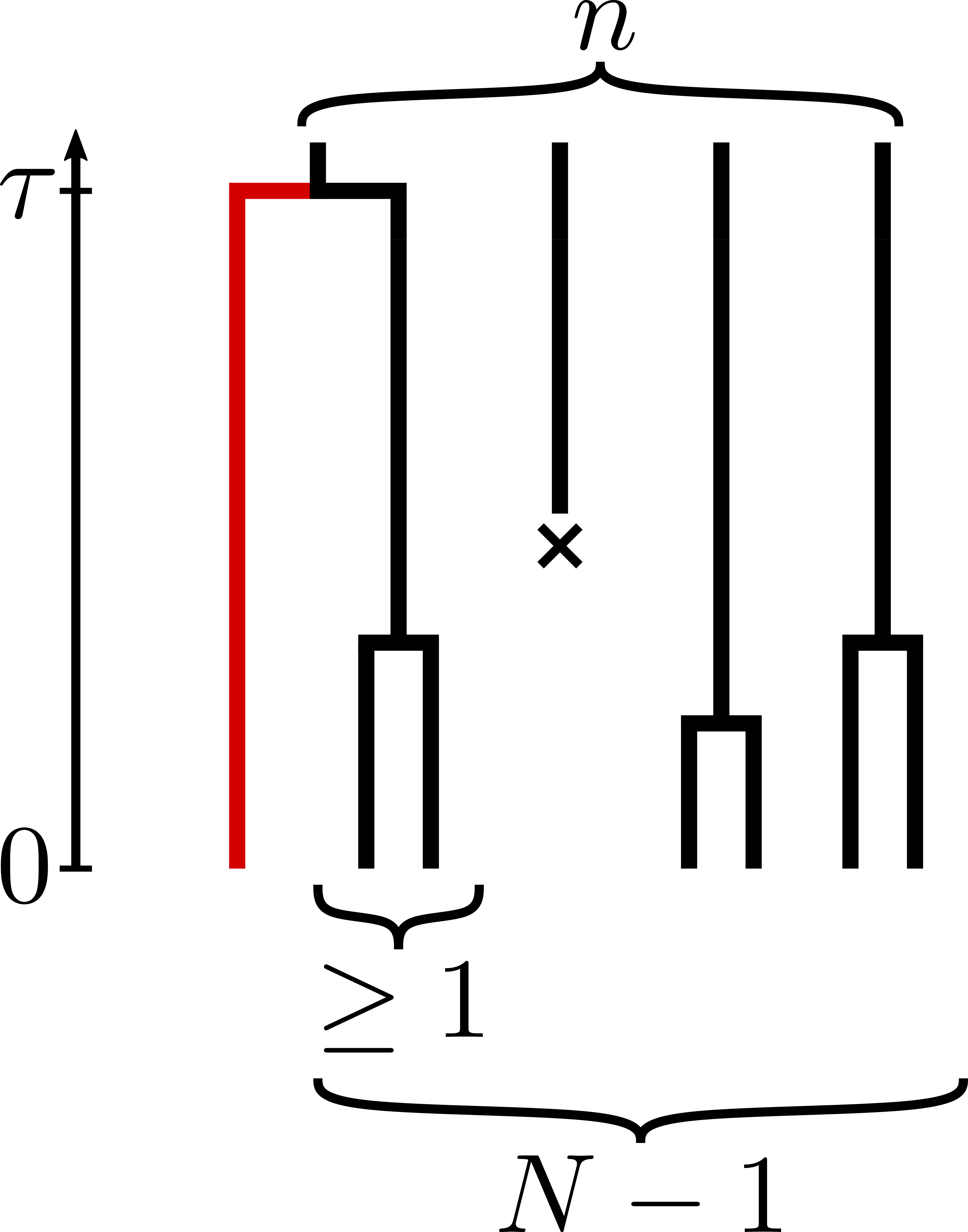} \hspace{0.1\textwidth}
		B\includegraphics[width=0.4\textwidth]{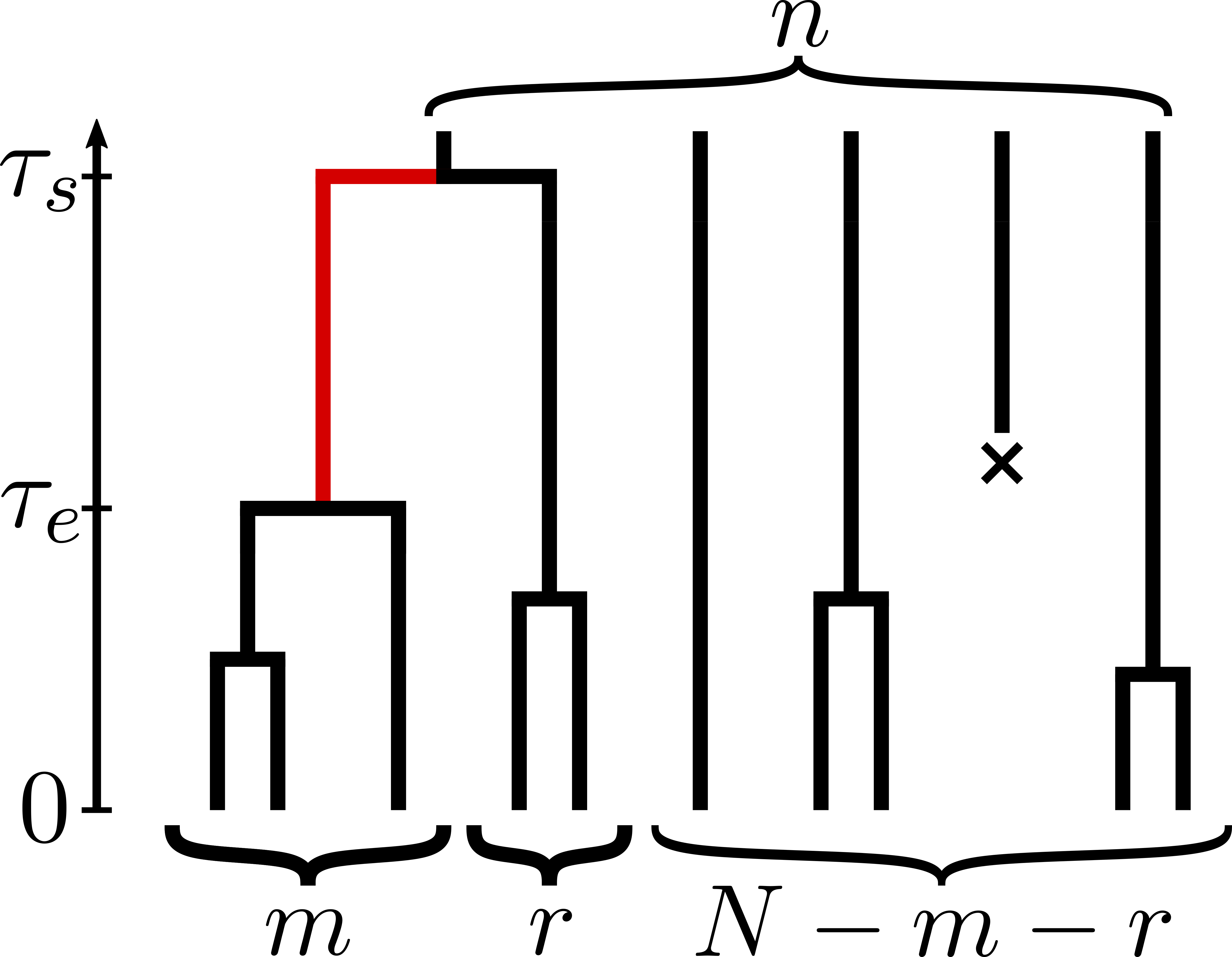}
		\caption{\textbf{Formation of branches given population size $N$}
			(A) Given a population size $N$ at the time of observation and $n$ individuals at time $\tau$ in the past, a pendant branch (red) forms from a birth event where one offspring has exactly one and the other offspring has at least one surviving descendant. The number of descendants from the latter and the other $n-1$ individuals at time $\tau$ must equal $N-1$. (B) The same situation, but for an interior branch: At time $\tau_s$ one of the $n$ individuals undergoes a birth event. To form an interior branch (red), one offspring has to form a lineage until time $\tau_e$ where all additional lineages arising die out before the population is observed. At time $\tau_e$, the individual undergoes a birth event with both offspring forming observed clades with a total of $m\geq2$ individuals. The second offspring from the birth event at $\tau_s$ cannot die out but forms a clade of size $r\geq1$. The remaining $N-m-r$ observed individuals have to be offspring of the $n-1$ individuals that did not divide at time $\tau_s$.
		}\label{fig:propagation}
	\end{figure}

	In the \new{first} approach, we look at all individuals just after the cell division event at time $\tau$ that establishes the pendant branch we are considering. Figure~\ref{fig:propagation}A sketches the situation: There are $n+1$ individuals, one of which has just established the pendant branch. This individual will have exactly one extant offspring at the time of observation. The other $n$ individuals must also have exactly one extant offspring in total \new{(which happens with probability $P(1\mid n)$)}, leading to a population size of $N=2$ at the time of observation. Additionally, the particular individual that was also born at time $\tau$ (but does not establish the branch we consider) must have at least one extant offspring. (For the case $N=2$, the branches starting with the birth event at time $\tau$ are equivalent, but these statements generalize to $N>2$).  The probability of the $n$ individuals having one extant offspring at the time of observation, with one given individual not going extinct is $P(1|n)-p_0 P(1|n-1)$\new{, where the latter term excludes the case where the $n$ individuals leave the desired single descendant, but the lineage of the second offspring of the birth event forming the pendant branch has died out.}
	In this approach, we rewrite the relative fraction of birth events leading to a pendant branch at time $\tau$ \new{(left-hand side of \eqref{eq:toyfinal})} as 
	\begin{align}
		\label{eq:appendix_secondapproach}
		\sum_{n=1}^{\infty}\lambda n p_1 \left(P(1|n)-p_0 P(1|n-1)\right) \new{\frac{p_{(n,2)}^{(\mu,\lambda)}}{ p_{(2,n)}^{(\lambda,\mu)}}}
		= 2 \lambda p_1 \sum_{n=1}^{\infty} \left(P(1|n)-p_0 P(1|n-1)\right)
		= 2 \lambda p_1 (1-p_0)\sum_{n=1}^{\infty} P(1|n) \ , 
	\end{align}
	which gives the same result as \new{before (equation~\eqref{eq:toyfinal})} since $P(1|n)=np_1 p_0^{n-1}$ and $\sum_{n=1}^{\infty} np_1 p_0^{n-1}=p_1 \partial_{p_0} \frac{p_0}{1-p_0}=\frac{p_1}{(1-p_0)^2}$. 
	
	The \new{second} approach uses a generating function to compute $P(1|n)$. \new{A generating function $g(x) \equiv \sum_n p_n x^n$ represents a probability distribution $p_n$ over a discrete set of outcomes as the coefficients of a power series. The coefficient $p_n$ can be recovered by taking the $n$-th derivative of $g$ with respect to $x$. The generating function can be interpreted as the expectation value of $x^n$, i.e.\ $g(x)=\mathrm{E}\left[x^n\right]$.} 

	The probability that a given clade has size $n$ at a time $\tau$ after it originated with a single individual is given by \eqref{eq:p0}-\eqref{eq:pn} and can be expressed as
	\begin{align}
		\label{eq:pn_app}
		p_n(\tau) &= \new{p_{(n,1)}^{(\lambda,\mu)}(\tau) = (1-p_0)\left(1-\frac{\lambda}{\mu}p_0\right) \left(\frac{\lambda}{\mu}p_0\right)^{n-1} = p_1 \left(\frac{\lambda}{\mu}p_0\right)^{n-1}} & n&>0 \nonumber \\
		p_0(\tau) &= \new{p_{(0,1)}^{(\lambda,\mu)}(\tau) = p_0} & n&=0  \ .
	\end{align}
	The corresponding generating function is
	\begin{equation}
		g(x) = \new{\mathrm{E}\left[x^n\right] =} \sum_{k=0}^\infty p_k(\tau)x^k = \new{p_0+(1-p_0)\left(1-\frac{\lambda}{\mu}p_0\right) \sum_{k=1}^\infty \left(\frac{\lambda}{\mu}p_0\right)^{k-1} x^k} \ .
	\end{equation}
	\new{This is an infinite geometric series. When starting with $n$ individuals, the total population size $N$ at time $\tau$ is described by the sum of the $n$ independent and identically distributed sizes $k_i$ of the clades generated by each individual, i.e.\ $N=\sum_{i=1}^n k_i$. Because $\mathrm{E}\left[x^N\right]=\mathrm{E}\left[x^k\right]^N$, the probability generating function of $N$ is simply the $n$th power of $g(x)$. Furthermore, }
	$P(1|n)$ is the coefficient of $x$ to the first power \new{in $g(x)^n$ and therefore} $P(1|n)=\partial_x \vert_{x=0} g(x)^n = np_1 p_0^{n-1}$ as in the second approach. However, since the geometric sum over $g(x)^n$ converges absolutely ($|\new{p_0}|<1$ and $\left|\new{\frac{\lambda}{\mu}p_0}\right|<1$ for $\mu<\lambda$ and $\tau>0$),
	the summation over $n$ and differentiation commute, so we can perform the sum over $n$ first
	\begin{equation}
		\sum_{n=1}^\infty P(1|n) = \sum_{n=1}^\infty \partial_x \vert_{x=0} g(x)^n=\partial_x\vert_{x=0} \sum_{n=1}^\infty g(x)^n=\partial_x\vert_{x=0} \frac{g(x)}{1-g(x)} \ .
	\end{equation}
	Straightforward algebra gives
	\begin{equation}
		\frac{g(x)}{1-g(x)} = \new{ \frac{p_0}{1-p_0} \frac{1}{1-x} + \frac{1-p_0-\frac{\lambda}{\mu}p_0}{1-p_0} \frac{x}{1-x}} \ .
	\end{equation}
	Since $\frac{x}{1-x}= -1+\frac{1}{1-x}$,  $\frac{x}{1-x}$ and $\frac{1}{1-x}$ have the same derivatives giving 
	\begin{equation}
		\sum_{n=1}^\infty P(1|n) =\partial_x\vert_{x=0} \new{\left[\frac{p_0}{1-p_0} \frac{1}{1-x} + \frac{1-p_0-\frac{\lambda}{\mu}p_0}{1-p_0} \frac{x}{1-x} \right]
		=\frac{1-\frac{\lambda}{\mu}p_0}{1-p_0}=\frac{p_1}{(1-p_0)^2} }
	\end{equation}
	as before. 
	
	The \new{second} approach based on the generating function generalizes immediately to any given final population size. With~\eqref{eq:relative_propagators}
	the relative fraction of birth events leading to a pendant branch is instead of \eqref{eq:appendix_secondapproach}
	\begin{align}
		N \lambda p_1 (1-p_0)\sum_{n=1}^{\infty} P(N-1|n)
	\end{align}
	and 
	\begin{align}
		\label{eq:app_sumPgivenn}
		\sum_{n=1}^{\infty} P(N-1|n)= \frac{1}{(N-1)!} \partial_x^{N-1}\vert_{x=0} \new{\left[\frac{p_0}{1-p_0} \frac{1}{1-x} + \frac{1-p_0-\frac{\lambda}{\mu}p_0}{1-p_0} \frac{x}{1-x} \right]
		=\frac{1-\frac{\lambda}{\mu}p_0}{1-p_0}=\frac{p_1}{(1-p_0)^2} } \ ,
	\end{align}
	independently of the final population size $N$. \new{Here, we have again used that the coefficient of the $(N-1)$th power of $x$ in $g(x)^n$ yields $P(N-1\mid n)$.} The resulting normalization factor turns out to be $2/N$, which means that the distribution of pendant branch lengths conditioned on the population size $N$ is independent of $N$ \new{and thus given by~\eqref{eq:toyfinal}}.
	
	Superficially, the result \eqref{eq:toyfinal} looks different from the distribution of pendant branch lengths derived in Stadler and Steel~\cite{stadler_distribution_2012} and Mooers et al.~\cite{mooers_branch_2012}, which in the present notation is $2\lambda p_{1}(\tau) \left(1-\frac{\lambda}{\mu} p_0(\tau)\right)$. The results agree, as they must, since \new{$\left(1-\frac{\lambda}{\mu} p_0(\tau)\right)=\frac{p_1(\tau)}{1-p_0(\tau)}$} (compare to \eqref{eq:pn_app}). Furthermore, the result coincides with the distribution of pendant branches conditioned on large tree ages $T$ \eqref{eq:ptau} derived in Section~\ref{sec:branches_pendant}, as $\new{\frac{p_1(\tau)}{(1-p_0(\tau))}}=\e^{-(\lambda-\mu)\tau}(1-p_0(\tau))$.
	
	\subsection{Interior branches}
	\label{app:cond_N_interior}
	The approach based on generating functions can also be applied straightforwardly to the statistics of interior branches conditioned on the final population size $N$. Figure~\ref{fig:propagation}B shows how a population of size $n$ at time $\tau_s$ gives rise 
	to an interior branch. The branch we focus on must have $m \geq 2$ extant offspring at the time of observation (with only one offspring it would be a pendant branch), the second individual born at time $\tau_s$ must have $r \geq 1$ offspring, and the remaining $n-1$ individuals have in total $N-r-m$ offspring. The latter condition ensures that the final population size equals $N$. The probability for such a trajectory starting from $n+1$ individuals at time $\tau_s$ and ending with $N$ at the time of observation is 
	\begin{align}
		p_{\mbox{mr}}(\tau_s) =P(N|n+1) - \left(p_0 P(N|n) +p_1 P(N-1|n)\right) - p_0 P(N|n) + \left( p_0 p_0 P(N|n-1) + p_1 p_0 P(N-1|n-1) \right) \ ,
	\end{align}
	where the suppressed time is $\tau_s$. 
	The first term gives the probability for ending up with $N$ individuals when starting with $n+1$, the next term in brackets excludes the two cases $m=0$ and $m=1$, and the term following that excludes the case $r=0$. The last term in brackets corrects overcounting the case when both $m<2$ \textit{and} $r=0$. 
	
	The next step looks at the clade establishing the interior branch at time $\tau_s$. Given that this clade has size $m \geq 2$ at the time of observation, what is the probability density (probability per small time period) of a birth event at time $\tau_e$, while the branch has no offspring with extant descendants prior to $\tau_e$ but is terminated at $\tau_e$, because both offspring establish extant lineages. This is 
	\begin{equation}
		\frac{p_1^{(\tau_e)}(\tau_s-\tau_e) \lambda \left(1-p_0(\tau_e)\right)^2}{1-p_0(\tau_s)-p_1(\tau_s)} \ ,
	\end{equation}
	where the denominator arises from conditioning on two extant or offspring $m \geq 2$. 
	
	Putting these results together, the probability density (up to a normalizing factor) at which interior branches start at time $\tau_s$ and end at time $\tau_e$ conditioned on a population size $N$ at the time of observation is 
	\begin{equation}
		\frac{p_1^{\tau_e}(\tau_s-\tau_e) \lambda \left(1-p_0(\tau_e)\right)^2 }{1-p_0(\tau_s)-p_1(\tau_s)} \sum_{n=1}^\infty \lambda n \new{\frac{p_{(n,N)}^{(\mu,\lambda)}(\tau_s)}{ p_{(N,n)}^{(\lambda,\mu)}(\tau_s)}} p_{\mbox{mr}(\tau_s)} \ .
	\end{equation}
	Although this expression looks cumbersome, it can be easily evaluated using~\eqref{eq:relative_propagators} and~\eqref{eq:app_sumPgivenn}.
	Using $P(N|0)=0$ and $\sum_{n=1}^\infty P(N|n+1)=\sum_{n=1}^\infty P(N|n)-P(N|1)$, the sum reduces to
	\begin{align}
		\sum_{n=1}^\infty p_{\mbox{mr}}(\tau_s) &= (1-2p_0(\tau_s)-p_1(\tau_s)+p_0(\tau_s)p_0(\tau_s)+p_1(\tau_s)p_0(\tau_s))\frac{p_1(\tau_s)}{(1-p_0(\tau_s))^2} - p_N(\tau_s) \nonumber \\
		&= (1-p_0(\tau_s)-p_1(\tau_s))\frac{p_1(\tau_s)}{1-p_0(\tau_s)} - p_N(\tau_s) 
		\label{eq:sum_p_mr} \ .
	\end{align}
	
	Thus, using equation \eqref{eq:pn_app} to replace $p_N(\tau_s)$, the probability density of an internal branch is proportional to
	\begin{align}
		\lambda^2 N p_1^{(\tau_e)}(\tau_s-\tau_e) \left(1-p_0(\tau_e)\right)^2 \left[ \frac{p_1(\tau_s)}{1-p_0(\tau_s)} - \frac{p_1(\tau_s)\left(\frac{\lambda}{\mu}p_0(\tau_s)\right)^{N-1}}{1-p_0(\tau_s)-p_1(\tau_s)} \right] \ .
	\end{align}
	The second term in the square brackets vanishes quickly with increasing population sizes. When neglecting this term, the normalizing factor again turns out to be $2/N$, removing the dependency on $N$ entirely. Hence, up to a term of order $\mathcal{O}(p_0(\tau_s)^{N-1})$, the distribution of interior branch lengths conditioned on the population size $N$ coincides with probability density function \eqref{eq:int_branch_propto}, the asymptotic distribution in trees conditioned on large $T$ as $p_1(\tau_s)/(1-p_0(\tau_s))=\e^{-(\lambda-\mu)\tau_s}(1-p_0(\tau_s))$. \\
	
	\subsection{Finite sampling}
	\label{app:cond_N_sampling}
	The results conditioned on a fixed final population size can be generalized to the finite sampling probability $\rho$ discussed in Section~\ref{sec:sampling}. Again, $p_0(\tau)$, $p_1(\tau)$ and $p_1^{(\tau_e)}(\tau)$ are replaced by $p_0^{(\rho)}(\tau)$, $p_1^{(\rho)}(\tau)$ and $p_1^{(\rho)(\tau_e)}(\tau)$ (equations \eqref{eq:p0_sampling}-\eqref{eq:p1_tau_e_sampling}). However, we do not know the population size $N$, only the number of samples.
	
	In the case of pendant branches, this means we do not know the exact number of extant descendants left by the $n$ individuals after the birth event at time $\tau$ who do not form the pendant branch (in the case of complete sampling they need to be $N-1$). However, since equation \eqref{eq:app_sumPgivenn} is independent of $N$, summing over all possible numbers of descendants leads to the same factor, and the density function of pendant branches is thus (up to normalization) given by equation \eqref{eq:toyfinal} with $p_0^{(\rho)}(\tau)$ and $p_1^{(\rho)}(\tau)$, coinciding with the result for trees at a given large age $T$ of the birth-death process in the case of finite sampling \eqref{eq:p_pend_f}.
	
	For the interior branch distribution, introducing finite sampling with $p_0^{(\rho)}(\tau)$, $p_1^{(\rho)}(\tau)$ and $p_1^{(\rho)(\tau_e)}(\tau)$ covers the asymptotic (and N-independent) case of large $N$, which again coincides with the probability density functions conditioned on large $T$ \eqref{eq:p_int_joint_f}-\eqref{eq:p_int_f}. The correction term for small $N$, however, is more challenging to generalize to finite sampling since $N$ is not known. In principle, it can be treated by summing over $N$ conditioned on a fixed number of samples and then following through with the steps in Appendix~\ref{app:cond_N_interior}.
	
\end{document}